\documentclass[draft]{agujournal2019}
\usepackage{url} %this package should fix any errors with URLs in refs.
\usepackage{lineno}
\usepackage[inline]{trackchanges} %for better track changes. finalnew option will compile document with changes incorporated.
\usepackage{soul}
\usepackage{xr}
\usepackage{array}

\usepackage{xcolor}
\usepackage{colortbl}

\journalname{JGR: Atmospheres}

\begin{document}

%% ------------------------------------------------------------------------ %%
%  Title
%
% (A title should be specific, informative, and brief. Use
% abbreviations only if they are defined in the abstract. Titles that
% start with general keywords then specific terms are optimized in
% searches)
%
%% ------------------------------------------------------------------------ %%

\title{Revisiting the definition of rapid intensification of tropical cyclones by clustering the initial intensity and inner-core size}

%% ------------------------------------------------------------------------ %%
%
%  AUTHORS AND AFFILIATIONS
%
%% ------------------------------------------------------------------------ %%

% Authors are individuals who have significantly contributed to the
% research and preparation of the article. Group authors are allowed, if
% each author in the group is separately identified in an appendix.)

% List authors by first name or initial followed by last name and
% separated by commas. Use \affil{} to number affiliations, and
% \thanks{} for author notes.
% Additional author notes should be indicated with \thanks{} (for
% example, for current addresses).

% Example: \authors{A. B. Author\affil{1}\thanks{Current address, Antartica}, B. C. Author\affil{2,3}, and D. E.
% Author\affil{3,4}\thanks{Also funded by Monsanto.}}

\authors{Yi Li\affil{1,2,3}, Youmin Tang\affil{3,4,5}, Ralf Toumi\affil{6}, Shuai Wang\affil{6,7}}

\affiliation{1}{Key Laboratory of Marine Hazards Forecasting, Ministry of Natural Resources, Hohai University, Nanjing, China}
\affiliation{2}{Key Laboratory of Ministry of Education for Coastal Disaster and Protection, Hohai University, Nanjing, China}
\affiliation{3}{College of Oceanography, Hohai University, Nanjing, China}
\affiliation{4}{University of Northern British Columbia, Prince George, Canada}
\affiliation{5}{Southern Marine Science and Engineering Guangdong Laboratory (Zhuhai), Zhuhai, China}
\affiliation{6}{Department of Physics, Imperial College London, London, UK}
\affiliation{7}{Atmospheric and Oceanic Sciences Program, Princeton University, Princeton, NJ, USA}
% \affiliation{4}{Fourth Affiliation}

% \affiliation{1}{=Affiliation Address=}
%(repeat as many times as is necessary)

%% Corresponding Author:
% Corresponding author mailing address and e-mail address:

% (include name and email addresses of the corresponding author.  More
% than one corresponding author is allowed in this LaTeX file and for
% publication; but only one corresponding author is allowed in our
% editorial system.)

% Example: \correspondingauthor{First and Last Name}{email@address.edu}

\correspondingauthor{Youmin Tang}{ytang@unbc.ca}

%% Keypoints, final entry on title page.

%  List up to three key points (at least one is required)
%  Key Points summarize the main points and conclusions of the article
%  Each must be 100 characters or less with no special characters or punctuation and must be complete sentences

% Example:
% \begin{keypoints}
% \item	List up to three key points (at least one is required)
% \item	Key Points summarize the main points and conclusions of the article
% \item	Each must be 100 characters or less with no special characters or punctuation and must be complete sentences
% \end{keypoints}

\begin{keypoints}
\item An objective definition of TC rapid intensification is developed by clustering the intensification rate, initial intensity, and size.
\item A new threshold of 45 kt/24h (23.2 m/s/24h), higher than the widely-used 30 kt/24h (15.4 m/s/24h), is determined for the ocean globally.
\item The thresholds for all basins are the same as the global value except for the North Atlantic and East Pacific where they are 40 kt/24h.
\end{keypoints}

%% ------------------------------------------------------------------------ %%
%
%  ABSTRACT and PLAIN LANGUAGE SUMMARY
%
% A good Abstract will begin with a short description of the problem
% being addressed, briefly describe the new data or analyses, then
% briefly states the main conclusion(s) and how they are supported and
% uncertainties.

% The Plain Language Summary should be written for a broad audience,
% including journalists and the science-interested public, that will not have 
% a background in your field.
%
% A Plain Language Summary is required in GRL, JGR: Planets, JGR: Biogeosciences,
% JGR: Oceans, G-Cubed, Reviews of Geophysics, and JAMES.
% see http://sharingscience.agu.org/creating-plain-language-summary/)
%
%% ------------------------------------------------------------------------ %%

%% \begin{abstract} starts the second page

\begin{abstract}
Rapid intensification (RI) of tropical cyclones (TCs) provides a great challenge in operational forecasting and contributes significantly to the development of major TCs.
RI is commonly defined as an increase in the maximum sustained surface wind speed beyond a certain threshold within 24 h.
The most widely used threshold is 30 kt (15.4 m/s), which was determined statistically.
Here we propose a new definition for RI by objectively clustering TCs using the intensification rate, initial intensity, and radius of the maximum wind speed.
A group of 770 samples is separated at a threshold of 45 kt (23.2 m/s). The threshold is 40 kt (20.6 m/s) for the western North Atlantic, where TC size measurements are more reliable.
Monte Carlo experiments demonstrate that the proposed threshold is robust even considering the uncertainty in RMW of as high as 30 km.
We show that, when a TC undergoes RI, its maximum wind speed is approximately 60 \textpm 15 kt (30.9 \textpm 7.7 m/s) and the radius of the maximum wind speed is 45 \textpm 20 km.
The new threshold outperforms the conventional threshold of 30 kt/24h in (1) describing the bimodal distribution of lifetime maximum intensity, and (2) explaining the annual count of Category 5 TCs. 
This new definition provides a more physically-based threshold and
describes a more reliable representation to the extreme events.
Although more comparisons are needed for operational application, it is likely to be desirable for process-based case studies, and could provide a more valuable metric for TC intensification classification and research.
\end{abstract}

% \section*{Plain Language Summary}

%% ------------------------------------------------------------------------ %%
%
%  TEXT
%
%% ------------------------------------------------------------------------ %%

%%% Suggested section heads:
% \section{Introduction}
%
% The main text should start with an introduction. Except for short
% manuscripts (such as comments and replies), the text should be divided
% into sections, each with its own heading.

% Headings should be sentence fragments and do not begin with a
% lowercase letter or number. Examples of good headings are:

% \section{Materials and Methods}
% Here is text on Materials and Methods.
%
% \subsection{A descriptive heading about methods}
% More about Methods.
%
% \section{Data} (Or section title might be a descriptive heading about data)
%
% \section{Results} (Or section title might be a descriptive heading about the
% results)
%
% \section{Conclusions}

\section{Introduction}
Rapid intensification (RI) is the dramatic strengthening of a tropical cyclone (TC) over a short period and poses a challenge for short-term weather forecasting and TC simulation
\cite{rappaport_joint_2012,rogers_noaas_2013,kaplan_evaluating_2015,cangialosi_recent_2020, demaria_operational_2021, wu_understanding_2022, emanuel_will_2017}. 
RI also has a substantial influence on the climatological distribution of TC intensity and most intense TCs undergo RI during their lifetimes \cite{lee_rapid_2016}.

Thus, RI is undoubtedly important in understanding TCs. 
RI is commonly defined as an increase in the maximum sustained surface wind speed (V$_{\mathrm{max}}$) of at least a certain threshold within 24 h.
The most widely used threshold was proposed by \citeA{kaplan_large-scale_2003}, who defined RI as the 95$^{\mathrm{th}}$ percentile of over-water 24-h intensity changes in the Atlantic TCs, which was 30 kt/24h.
However, many other thresholds exist in the literature. 
For instance, \citeA{kaplan_revised_2010} discussed the thresholds of 25 (12.9 m/s), 30 (15.4 m/s) and 35 kt (18.0 m/s) per 24 h, and highlighted the practical importance of the RI threshold for operational forecasting. \citeA{lee_rapid_2016} documented 35 kt/24h as the optimal RI threshold describing the bimodal distribution of the lifetime maximum intensity (LMI).

% Therefore, one may reasonably question whether these statistical thresholds are robustly defined despite their widespread use. 
The above statistical thresholds are invaluable and have been widely used in analyses of both internal and environmental processes during RI.
% The conventional logic is to regard RI as an extreme event, and then study the processes related to it. 
% However, one may reasonably question whether thresholds can be defined other than statistically.
However, the conclusions depend heavily on the chosen threshold, and one may reasonably question whether thresholds can be defined by other ways rather than statistically.
A more rigorous definition of RI is desirable, especially for process-based studies.
Here we revised the conventional threshold via clustering intensification-related metrics, and thereby proposed a threshold considering some of physical properties of the vortex. 
The physical mechanisms underlying TC intensification are complex and much effort has been devoted to understanding them \cite{wang_current_2004, montgomery_paradigms_2014, emanuel_100_2018}. 
Environmental and inner-core processes may both play important roles in determining whether and when a TC will undergo RI. 
Here, we focus on the initial vortex state, represented by TC intensity and vortex size metrics, such as V$_{\mathrm{max}}$ and the radii of gale-force (34-kt, R$_{34}$) and maximum winds (RMW). These metrics were selected because they are routinely reported by the operational centers and their effects on intensification have been demonstrated in both observational and theoretical studies. 

A TC of medium intensity is more likely to undergo RI \cite{kaplan_revised_2010, xu_dependence_2018, wang_intensity-dependence_2021}.
When a TC is relatively weak, the inner-core inertial stability and heating efficiency increase with intensity, and thus, the TC intensifies \cite{shapiro_response_1982,schubert_inertial_1982}.
However, as the intensity of the storm nears its maximum potential intensity, the frictional dissipation counteracts the heating efficiency, and the intensification rate decreases.
The maximum intensification rate is typically observed at approximately 60$\sim$70 kt, where the intensification potential is close to the weakening rate due to surface friction \cite{kaplan_revised_2010,wang_intensity-dependence_2021}.

In addition, the vortex size reflects feedback between the primary circulation and inner-core processes \cite{mallen_reexamining_2005}, and affects the subsequent intensification \cite{emanuel_finite-amplitude_1989}. 
% A small RMW favors , 
Studies based on best-track data have confirmed that RI events rarely occur when RMW is larger than 100 km, because a small RMW favors intensification due to the conservation of angular momentum \cite{carrasco_influence_2014, xu_dependence_2018}.
% A small RMW also favors intensification due to the conservation of angular momentum.
% Both numerical and observational studies \cite{li_why_2021,li_how_2022,wang_outer_2020,sitkowski_intensity_2011,fischer_rapid_2020} showed the typical RMW during RI is approximately 30 to 50 km, which is consistence with our results.
It has also been known that the inward contraction of eye-wall, usually represented by a decrease in RMW, often occurs simultaneously with intensification \cite{shapiro_response_1982,schubert_inertial_1982}.
% For the perspective of conservation of angular momentum, \citeA{li_why_2021}
In a seminal study, \citeA{shapiro_response_1982} showed that the response of low-level tangential wind tendency to diabatic heating is greater inside RMW than outside.
Thus, when RMW decreases, the potential energy of the vortex increases and the balanced tangential winds intensify simultaneously. 
Recent studies also show that rapid contraction of RMW could occur prior to RI \cite{stern_revisiting_2015, li_revisiting_2019,li_why_2021,li_how_2022, wu_rapid_2021}.
However, cases also exist in which RI co-occurs with steady RMW \cite{kieu_investigation_2012,qin_rapid_2018}.
Therefore, we only considered RMW instead of its changes in this work.

Other metrics that reflect the outer size of the TC, specifically the average radius of the 34-kt wind speed (R34) and TC fullness, were also examined in this study. 
Notably, \citeA{carrasco_influence_2014} showed that R$_{34}$ aids in RI prediction. 
In addition, \citeA{guo_tropical_2017} proposed the concept of TC fullness, a metric calculated from RMW and R$_{34}$ [defined as $(1 - \frac{\mathrm{RMW}}{\mathrm{R}_{34}})$], and demonstrated that a high TC fullness is essential for intense TC development. 
%Inter-basin difference: Inter-basin difference has been reported. 
In this study, a joint clustering method was implemented based on the K-means algorithm.
Such clustering algorithms have been widely used in TC research. 
For instance, \citeA{camargo_cluster_2007-1} utilized joint clustering analysis to group western North Pacific TC tracks based on their locations of genesis and subsequent tracks. 
\citeA{arnott_characterization_2004} analyzed the characteristics of extra-tropical transitions based on the results of K-means. 
\citeA{guo_tropical_2017} clustered the evolution curves of the Atlantic Hurricanes to analyze TC fullness.
Moreover, clustering has been widely used to detect outliers or extreme events \cite<e.g.,>{chawla_k_2013}. 
Therefore, we chose to detect extreme intensification using this technique.
Using a joint clustering algorithm, the common features for the RI events were extracted and the RI and non-RI clusters were subsequently separated.
We focused on Atlantic TCs to the west of 55$^{\circ}$W from 2004 to 2020, mainly considering better data quality than those for other basins and periods.
Our data pre-processing and main methods, including the unsupervised clustering technique, are introduced in Section 2. The sensitivity experiments and statistical characteristics of TCs undergoing RI, as defined with the new threshold, are presented in Section 3, where the performance of this new threshold is also explored. Section 4 discusses the main results and summarizes the key findings.

\section{Data and Methods}
TC data were obtained from the International Best Track Archive for Climate Stewardship \cite<IBTrACS, v4r00,>{knapp_international_2010}. 
For consistency and quality, we only utilized data from the National Hurricane Center for the Atlantic and East Pacific, and from the Joint Typhoon Warning Center for the remainder of the globe, for 2004–2020. 
The two USA agencies measure intensity as the 1-min sustained maximum 10-m wind speed (V$_{\mathrm{max}}$).
The intensification rate was defined as the change of V$_{\mathrm{max}}$ during each 24-h interval (hereafter $\Delta$V$_{24}$). 
We only obtained records for the standard observational times: 00, 06, 12, and 18 Coordinated Universal Time (UTC). 
RI may occur for consecutive 24-h periods and some of the tracks overlapped. 
In our study, we only selected TCs between 30$^{\circ}$N and 30$^{\circ}$S to minimize the influence of extra-tropical transition. To eliminate the topographic effects, we chose only TC tracks over the ocean, and all TC centers were at least 100 km from the coastline. The distance to the nearest landmass was provided by IBTrACS for each TC location. 
% We only selected TCs whose centers were at least 100 km away from coast and between 30$^{\circ}$N and 30$^{\circ}$S.
% The distance to the nearest landmass is provided by the IBTrACS for each TC location.
This pre-processing is similar to previous studies \cite<e.g.,>{kaplan_large-scale_2003,ma_definition_2019}.

Environmental factors were also analyzed, with the relative humidity, wind, and sea surface temperature (SST) data obtained from the fifth generation of atmospheric reanalysis from the fifth version of ECMWF re-analysis \cite<ERA5,>{hersbach_era5_2020}. 
The horizontal and temporal resolutions of the ERA5 data are 0.25$^{\circ}$ and 6 hours, respectively.
For ERA5 we also analyzed the data for 00, 06, 12, and 18 only.

Since 2004, the wide use of satellite, in situ observations, aircraft reconnaissance and routine post-season analyses have significantly reduced the observation error, especially for the Atlantic basin to the west of 55$^{\circ}$W.
Unfortunately, large uncertainties still exist in the RMW data. 
Even for western North Atlantic, RMW was not best tracked until the 2021 hurricane season \cite{landsea_revised_2022}, although the data during earlier periods has been utilized extensively for various basins \cite<e.g.,>{xu_statistical_2015, xu_dependence_2018, wu_rapid_2021, li_how_2022}.
Thus an independent test was performed, by applying the best-tracked data of 2021 to the clustering model that was trained with the data from 2004 to 2020. 
\citeA{landsea_revised_2022} also reported that the uncertainty with RMW observation can be as high as 16 nautical miles (30 km) for TCs stronger than 64kt (Category 1). 
As in \citeA{li_rapid_2022}, we carried out Monte Carlo experiments here to estimate the potential influence of observation uncertainties, in which 1,000 samples were produced by adding random noise from a Gaussian distribution with a mean of 0 and a standard deviation of 30 km to each RMW value.
In addition, to estimate the influence of uncertainty from V$_{\mathrm{max}}$ measurements, we conducted another Monte Carlo experiment, in which 1,000 subsamples were produced by adding random noise from a uniform distribution on the interval \textpm10 knots to each intensity change value \cite{bhatia_recent_2019}.

% we also compared the western North Atlantic and all other basins. 
% The intensity data is reliable over this period due to the wide use of satellite observations and the adjustment of the algorithms regarding the shift of wind-pressure relationships \cite{Kossin2018ASpeed}. 

The clustering algorithm requires input samples, which consisted of a combination of the initial V$_{\mathrm{max}}$ and/or initial TC size metrics (RMW, R$_{34}$, TC fullness) at t = 0 h, and the $\Delta$V$_{24}$ of the subsequent 24-h period. 
The primary clustering algorithm used in this study was K-means. 
This algorithm partitions the samples into subsets while minimizing the variance within each cluster.
In addition, another unsupervised artificial neural network, the self-organizing map (SOM) \cite{kohonen_self-organizing_1990, kohonen_essentials_2013, liu_performance_2006} was used for comparison. The SOM fits initially random weights to observations by comparing their Euclidean distances through competitive training. 
For both methods, the number of clusters, $K$, significantly affects the clustering results; their performance can be evaluated using the silhouette \cite{rousseeuw_silhouettes_1987} and Davies-Bouldin scores \cite{davies_cluster_1979}. The silhouette score measures the similarity of the points within each cluster and is defined as follows:

 \begin{linenomath*}
 \begin{equation} \label{eq_silhouette}
S = \frac{b-a}{\mathrm{max}(a,b)},
 \end{equation} 
 \end{linenomath*}
 
\noindent where $a$ is the mean distance between one point and all other points in the same cluster and $b$ is that for the nearest cluster. Thus a higher silhouette score indicates higher similarity within each cluster and a better clustering model. In contrast, the Davies-Bouldin score compares the average similarity between each pair of clusters. It can be defined as follows: 

 \begin{linenomath*}
 \begin{equation} \label{eq_db}
DB = \frac{1}{k} \displaystyle\sum_{i=1}^{k} \max_{i \neq j} R_{ij},
 \end{equation} 
 \end{linenomath*}

\noindent where $R_{ij}$ measures the ratio between cluster diameter ($s_i$, i.e., the average distance between cluster centroid and the points within the cluster $C_i$) and the distance between the currents $C_i$ and $C_j$. Therefore, a smaller Davies-Bouldin score indicates a better clustering result. Both metrics were used in this study to calculate the distance of clusters in the parameter space and determine the optimal $K$.

The V$_{\mathrm{max}}$, RMW, R$_{34}$ and TC fullness data were used for clustering. The environmental factors were not analyzed for clustering because only the properties of the initial vortex were considered in this study.
The intensity and size data were pre-processed and normalized before input to the clustering models, because their units and scales are different. All the variables are available for each input sample. It should be noted that we only consider the intensifying events for clustering due to their practical importance. 
Seven sensitivity experiments were performed to find the optimal combination of input variables, and the main model we used here clustered the input of V$_{\mathrm{max}}$, RMW and $\Delta$V$_{24}$ into 8 clusters.
The details of these experiments are summarized in the supplementary material (Text S1 and Table S1). 

% Sensitivity experiments were performed to test different combinations of input variables, as summarized in Table~\ref{tab_exp}:

% \begin{table}
% \caption{The input variables for .}  \label{tab_exp}
% \centering
% \begin{tabular}{c c }
% \hline
%   &  Input Variables  \\
% \hline
%   Exp\_V & Initial V$_{\mathrm{max}}$, $\Delta$V$_{24}$ \\
%   Exp\_V\_RM & Initial V$_{\mathrm{max}}$, RMW, $\Delta$V$_{24}$ \\
%   Exp\_V\_RM & Initial V$_{\mathrm{max}}$, RMW, $\Delta$V$_{24}$ \\
%   Exp\_V\_R$_{34}$ & Initial V$_{\mathrm{max}}$, RMW \\
%   Exp\_V\_TCF & Initial V$_{\mathrm{max}}$, RMW \\
%   Exp\_V\_RM & Initial V$_{\mathrm{max}}$, RMW \\
%   Exp\_V\_TCF & Initial V$_{\mathrm{max}}$, RMW \\
% \hline
% % \multicolumn{5}{l}{}
% \end{tabular}
% \end{table}

% the initial V$_{\mathrm{max}}$ and $\Delta$V$_{24}$ (hereafter V\_DV); initial V$_{\mathrm{max}}$, RMW and $\Delta$V$_{24}$ (hereafter V\_RMW\_DV); initial RMW and $\Delta$V$_{24}$ (hereafter RMW\_DV); initial V$_{\mathrm{max}}$, TC fullness [defined as $(1 - \frac{\mathrm{RMW}}{\mathrm{R}_{34}})$], and $\Delta$V$_{24}$ (hereafter V\_TCF\_DV); and initial V$_{\mathrm{max}}$, R$_{34}$, and $\Delta$V$_{24}$ (hereafter V\_R$_{34}$\_DV).
% \noindent The intensity and size data were pre-processed and normalized before input to the clustering models, because their units and scales are different. All the variables are available for each input sample. It should be noted that we only consider the intensifying events for clustering due to their practical importance. 

\section{Results}
\subsection{Overall distribution of the intensification rate}
\begin{figure}[ht!]
\centerline{\includegraphics[height=4in]{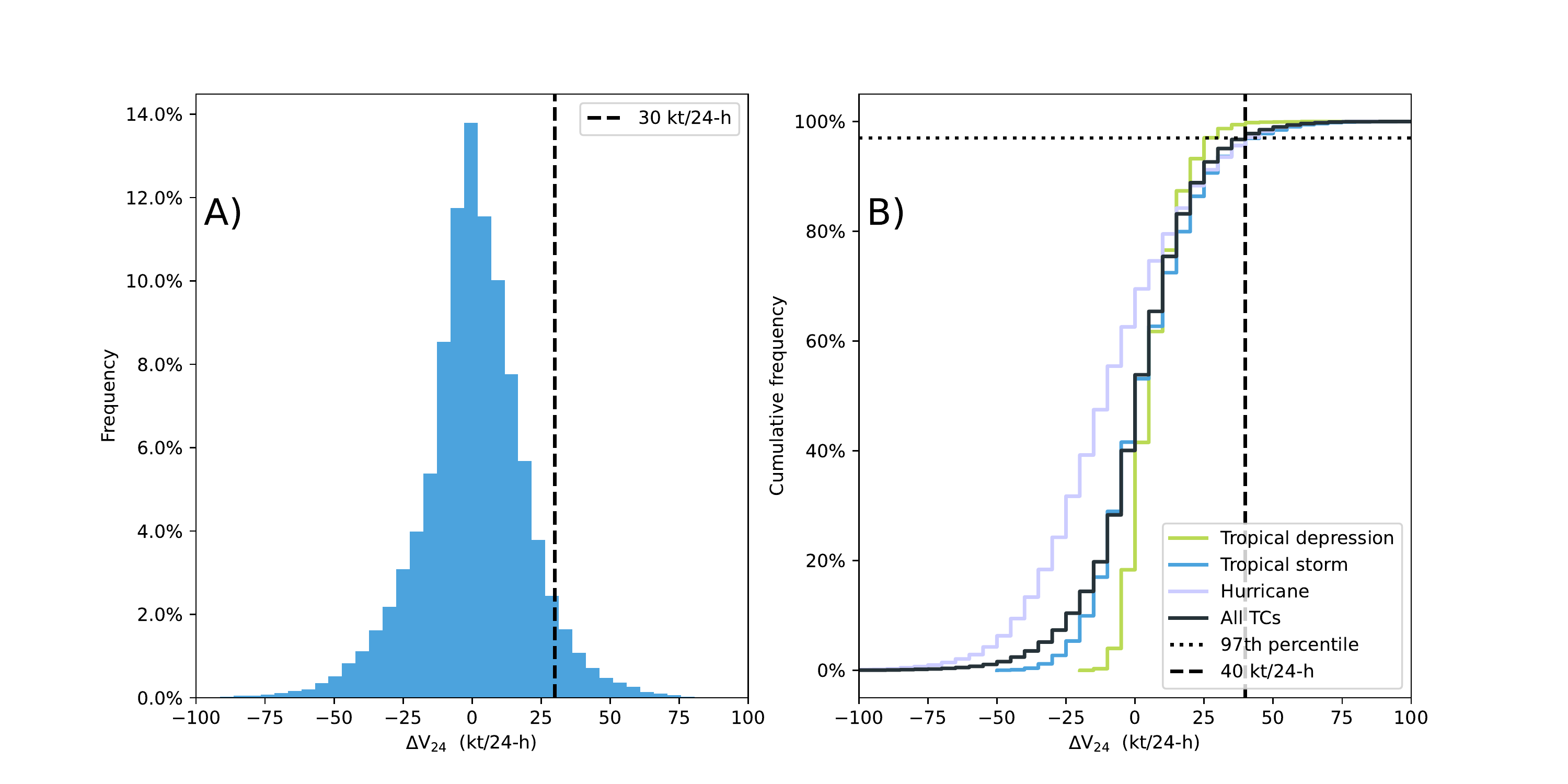}}
\caption{(A) The frequency distributions of 24-h intensification rate ($\Delta$V$_{24}$); (B) The cumulative frequency distributions of $\Delta$V$_{24}$ for the initial intensity of different scales, i.e., Tropical depressions (V$_{\mathrm{max}} \leq 33$ kt), Tropical storms ($34 \leq$ V$_{\mathrm{max}} \leq 63$ kt) and Hurricanes (V$_{\mathrm{max}} \geq 64$ kt). Note the distributions is this figure were calculated using all $\Delta$V$_{24}$ instead of the positive values alone.} \label{fig_CDF}
\end{figure}

A total of 886 storms and 12530 intensifying events were extracted from the global best-track data, and 107 storms and 1443 events for the western North Atlantic basin. The probabilistic distribution (Fig.~\ref{fig_CDF}A) of $\Delta$V$_{24}$ for the global basins is continuous and no discernible gap appears near 30 kt/24h, which is the traditional threshold. 
Moreover, this continuity suggests that the traditional threshold, which only considers the intensification rate, is a purely statistical or practical choice. This is similar to the findings of \citeA{kowch_are_2015}.
Fig.~\ref{fig_CDF}B 
depicts the cumulative frequency distribution, similar to that in Fig. 2 of \citeA{kaplan_large-scale_2003}, but using our data for all TCs instead of the North Atlantic alone. However, the curves for the different initial intensities did not converge until $\Delta$V$_{24}$ was above the 97$^{\mathrm{th}}$ percentile (approximately 40 kt/24h). This discrepancy partially justifies the need for further examination of the statistical RI thresholds. 

\begin{figure}[ht!]
\centerline{\includegraphics[width=8in]{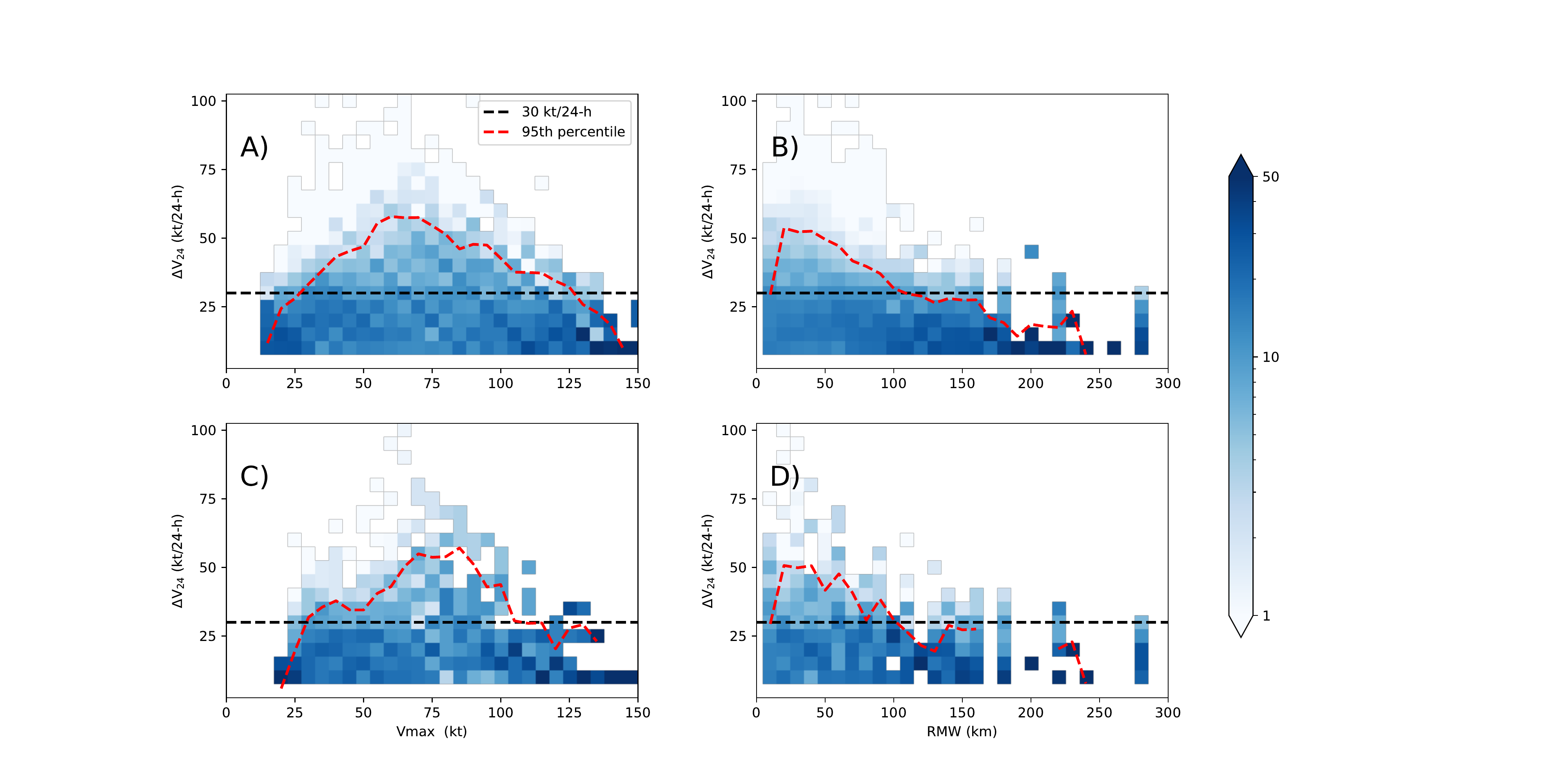}}
\caption{The ratio of positive $\Delta$V$_{24}$ as a function of (A) the initial maximum sustained surface wind speed (V$_{\mathrm{max}}$) and (B) the initial radius of the maximum wind (RMW) for the global basins. (C) and (D) are the same as (A) and (B) but for the western North Atlantic basin. The y-axis in each subplot is $\Delta$V$_{24}$ (units: kt/24h) and the x-axis shows the initial V$_{\mathrm{max}}$ (units: kt) and RMW (units: km), respectively. The unit for the blue shadings is \%.} \label{fig_den}
\end{figure}

Intensification is likely strongly influenced by initial state of the vortex, which are reflected in part by the initial intensity and size.
Thus, it is useful to analyze the joint frequency distributions of $\Delta$V$_{24}$ and the initial V$_{\mathrm{max}}$ and RMW (Fig.~\ref{fig_den}).
The $\Delta$V$_{24}$ increased with V$_{\mathrm{max}}$ when V$_{\mathrm{max}}$ was lower than approximately 60 kt (30.9 m/s), and $\Delta$V$_{24}$ could be up to 100 kt/24h. When the TC was stronger than 75 kt (38.6 m/s), $\Delta$V$_{24}$ decreased. TCs with this intensity range are generally well organized and far from their maximum potential intensity \cite{kaplan_revised_2010, wang_intensity-dependence_2021}. 
Additionally, RI was also more likely to arise when RMW is less than 100 km, and $\Delta$V$_{24}$ generally decreased as the RMW increased. A smaller RMW reflects higher inner-core inertial stability and dynamic efficiency, which enhances intensification  \cite{schubert_inertial_1982}. 
The 95$^{\mathrm{th}}$ percentile of $\Delta$V$_{24}$ is also significantly higher in this range of RMW and V$_{\mathrm{max}}$. 
For the North Atlantic, where the TC size measurement is more reliable, $\Delta$V$_{24}$ was generally lower; however, the V$_{\mathrm{max}}$ and RMW trends were similar. $\Delta$V$_{24}$ increased with V$_{\mathrm{max}}$ at values of less than approximately 60 kt, and decreased with V$_{\mathrm{max}}$ at values of higher than 80 kt. RI generally occurred when RMW was smaller than 100 km. 
These results align with the findings from previous studies \cite{carrasco_influence_2014,xu_relationship_2016,xu_dependence_2018}. 
In addition, RI mainly occurs with medium V$_{\mathrm{max}}$ and small RMW, which improves the separation of RI events in parameter space and promotes clustering performance.

\subsection{Sensitivity experiments}
\begin{figure}[ht!]
\centerline{\includegraphics[height=4in]{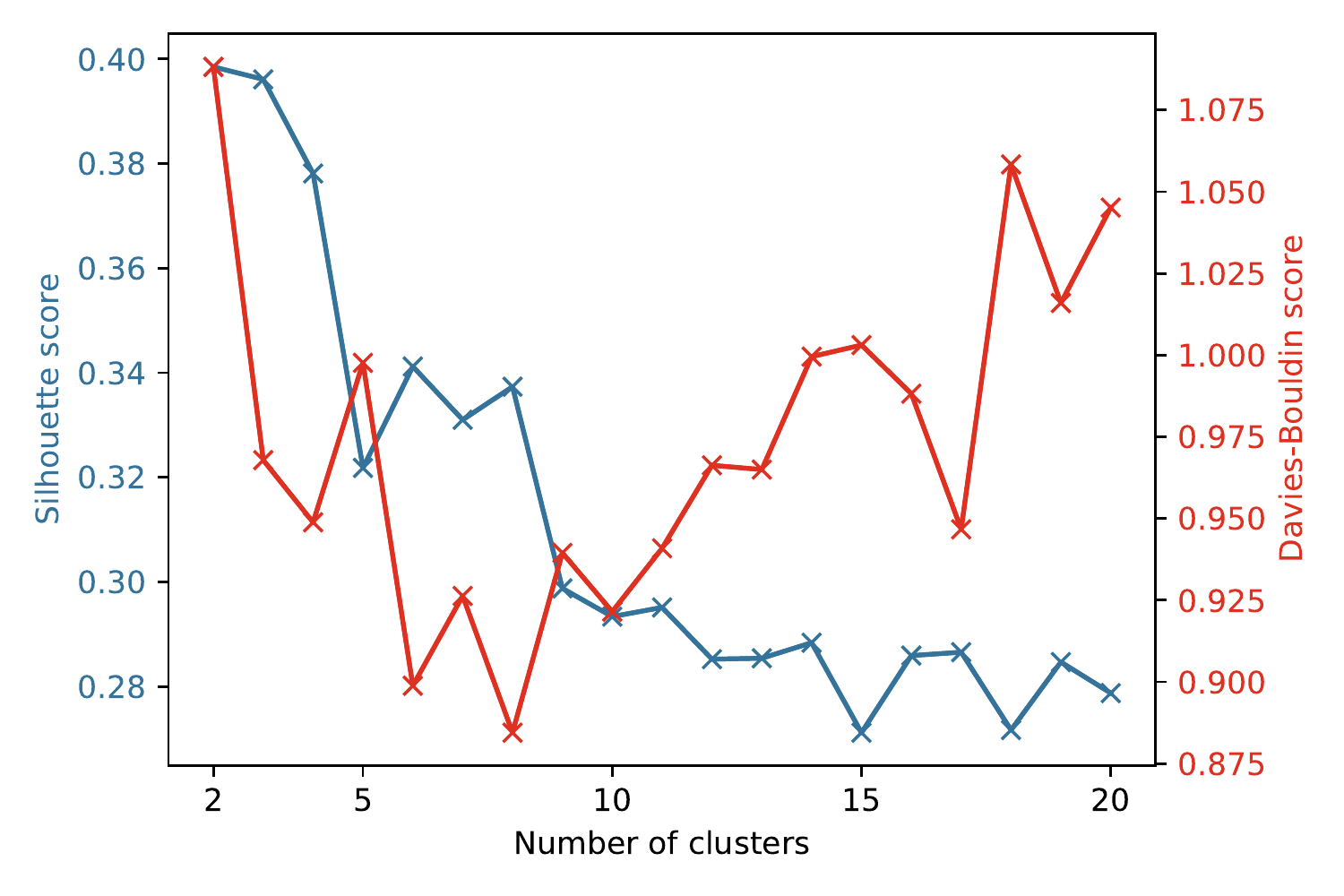}}
\caption{The silhouette and Davies-Bouldin scores as functions of the number of clusters when initial V$_{\mathrm{max}}$, RMW and $\Delta$V$_{24}$ are input for clustering. The models were trained using global data.} \label{fig_score}
\end{figure}

\begin{figure}[ht!]
\centerline{\includegraphics[width=8in]{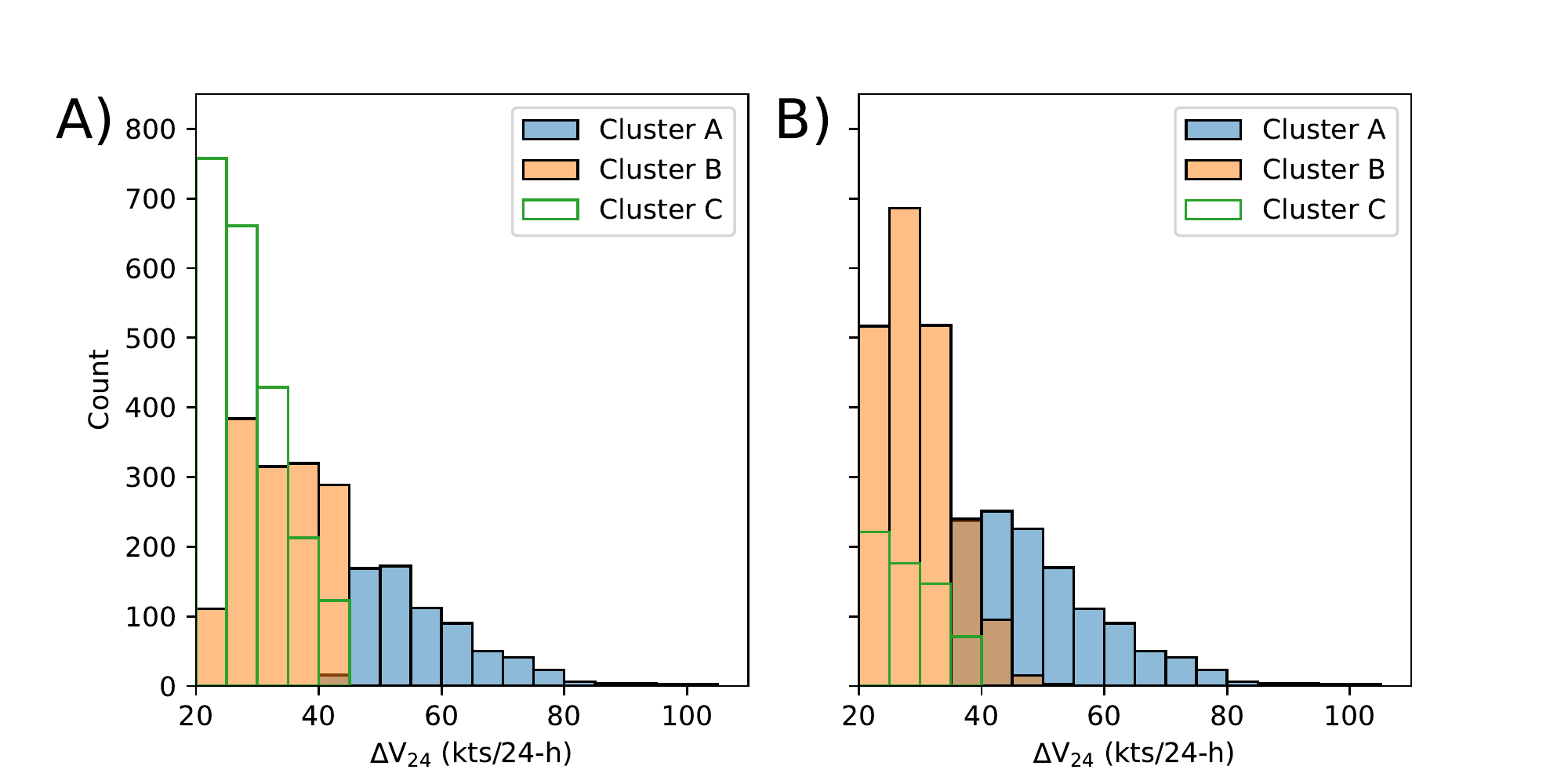}}
\caption{Intensification rate ($\Delta$V$_{24}$) of different clusters with (A) eight and (B) six clusters when initial V$_{\mathrm{max}}$, RMW and $\Delta$V$_{24}$ are input for clustering. The rapid intensification (RI) cluster is labeled as Cluster A in both subplots. The models were trained using global data. Only the three most rapidly intensifying clusters are plotted for better display clarity. The results with all clusters can be found in Fig. S1.} \label{fig_dv_vm_rm_3}
\end{figure}

We performed a series of sensitivity experiments to choose an optimal number of clusters ($K$) and combination of input variables. 
The best performance is found in the model with initial V$_{\mathrm{max}}$, RMW and $\Delta$V$_{24}$ as input, which is shown here as an example.
When the intensifying events were clustered into 8 groups, the Davies-Bouldin score reached a minimum, and the silhouette score was relatively high (Fig.~\ref{fig_score}).
These metrics indicated the distance between clusters reaches the minimum in the three-dimensional space (V$_{\mathrm{max}}$, RMW and $\Delta$V$_{24}$) when clustered into 8 groups.
We then analyzed $\Delta$V$_{24}$ with such a configuration, and an 'RI' cluster with a significantly higher $\Delta$V$_{24}$ was distinctly separated from the remainder of the dataset (Fig.~\ref{fig_dv_vm_rm_3}A).
The minimum $\Delta$V$_{24}$ of this 'RI' cluster is 45 kt/24h.
Fig.~\ref{fig_score} suggests that clustering the TCs into six groups was also reasonable.
However, the overlaps between the RI and non-RI clusters were much broader, although RI clusters still emerged at a threshold of approximately 35-40 kt/24h (Fig.~\ref{fig_dv_vm_rm_3}B). Therefore, the 8-cluster model was chosen owing to its superior overall performance. Most of the 770 TC events in the RI cluster had a $\Delta$V$_{24}$ greater than or equal to 45 kt/24h. 
For North Atlantic, on the other hand, a similar optimal configuration was also found, and the threshold is 40 kt/24h for North Atlantic.
The detailed results of other sensitivity experiments can be found in the supplementary text.

% The results show that, clustering IR with both RMW and V$_{\mathrm{max}}$ into 8 clusters is the optimal choice, such that the Davies-Bouldin score reaches the minimum and the Silhouette score is relatively high, and the RI and non-RI clusters are robustly and clearly distinguished. 
% The details of the sensitivity experiments are described in the supplementary materials.

\begin{figure}[ht!]
\centerline{\includegraphics[width=8in]{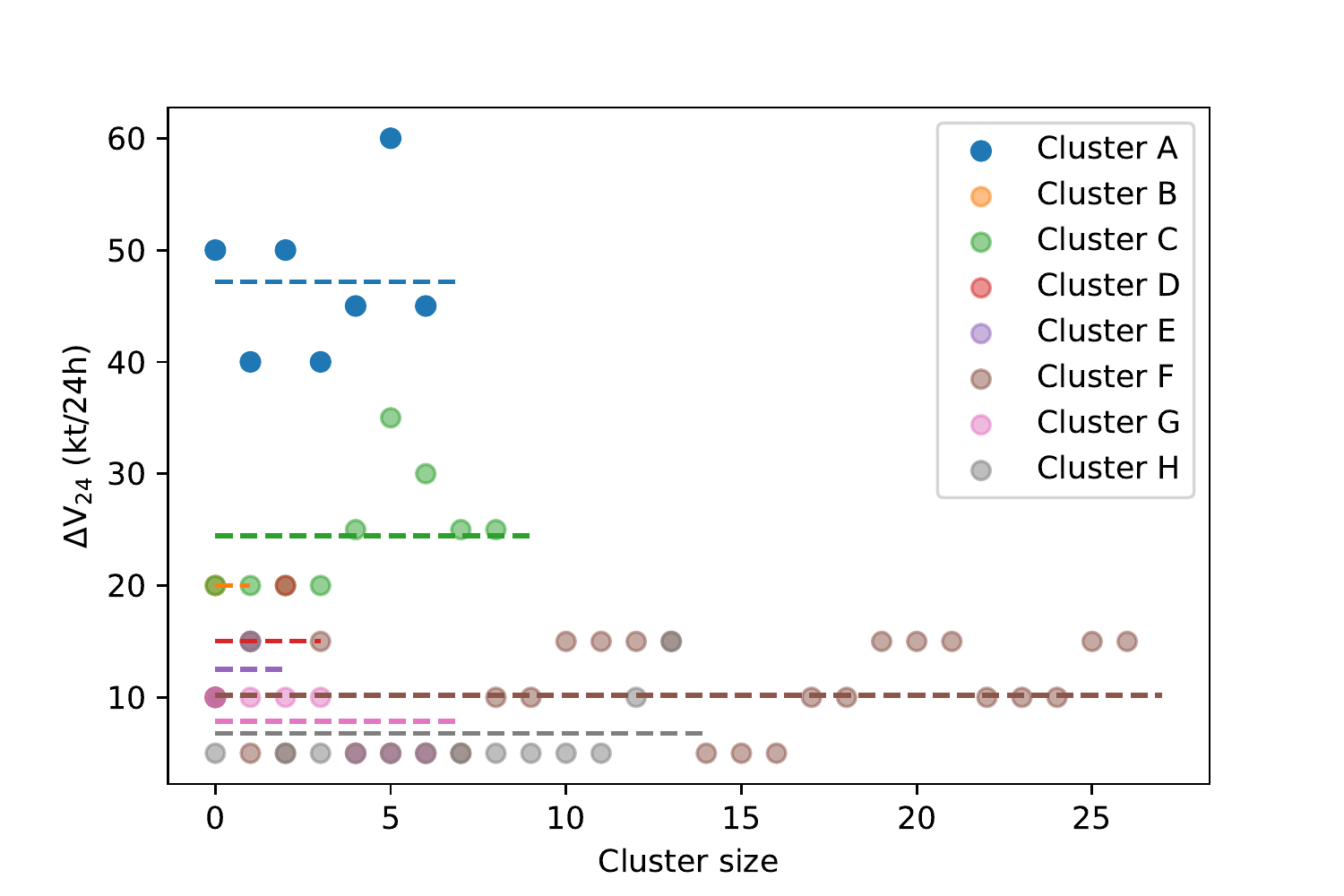}}
\caption{Intensification rate ($\Delta$V$_{24}$) of different clusters for the 2021 North Atlantic hurricane season. The rapid intensification (RI) cluster is labeled as Cluster A. The dashed lines depict the average $\Delta$V$_{24}$ for each cluster.} \label{fig_2021}
\end{figure}

% With the above clustering model configuration, an 'RI' cluster was separated for the global basin and the minimum $\Delta$V$_{24}$ is 45 kt/24h.
% For North Atlantic, the threshold is 40 kt/24h.
% However, it is noteworthy that the robustness of above results highly depend on the quality of best-track data, especially considering the uncertainty with RMW is high. 
The robustness of clustering highly depends on the quality of best-track data, especially considering the uncertainty with RMW is high.
In the Monte Carlo experiments, we found the threshold of 45 kt/24h is robust even when the uncertainty of RMW was set to 30 km.
In the 1,000 perturbed samples, 95\% (952) achieved the same threshold of 45 kt/24h while 5\% (48) produced a threshold of 40 kt/24h.
In addition, when the North Atlantic model was applied to the 2021 hurricane season, an RI group was also classified and the minimum $\Delta$V$_{24}$ (i.e., threshold) was 40 kt/24h (Fig.~\ref{fig_2021}), consistent with the threshold identified using data of earlier periods.
% An RI group was also separated and the minimum V$_{\mathrm{max}}$ was 40 kt/24h (Fig.~\ref{fig_2021}).
The results of above two tests indicate the proposed threshold is robust, even considering the high uncertainty of RMW measurements.
Nevertheless, the threshold of 30 or 35 kt/24h was not found in these experiments.

\subsection{Characteristics and evaluation of the new threshold}
The threshold of 45 kt/24h was selected due to its better overall performance in the sensitivity experiments. 
This threshold corresponds to the 97$^{\mathrm{th}}$ percentile of the global over-water $\Delta$V$_{24}$.
For the global basins, the mean $\Delta$V$_{24}$, V$_{\mathrm{max}}$ and RMW of this cluster were 54 kt/24h, 62 kt and 45 km, respectively (Table~\ref{tab_ri}). 
On the other hand, these properties for the TCs with a $\Delta$V$_{24}$ of over 30 kt/24h, were 40 kt/24h, 58 kt and 52 km, respectively. 

\begin{figure}[ht!]
\centerline{\includegraphics[width=6in]{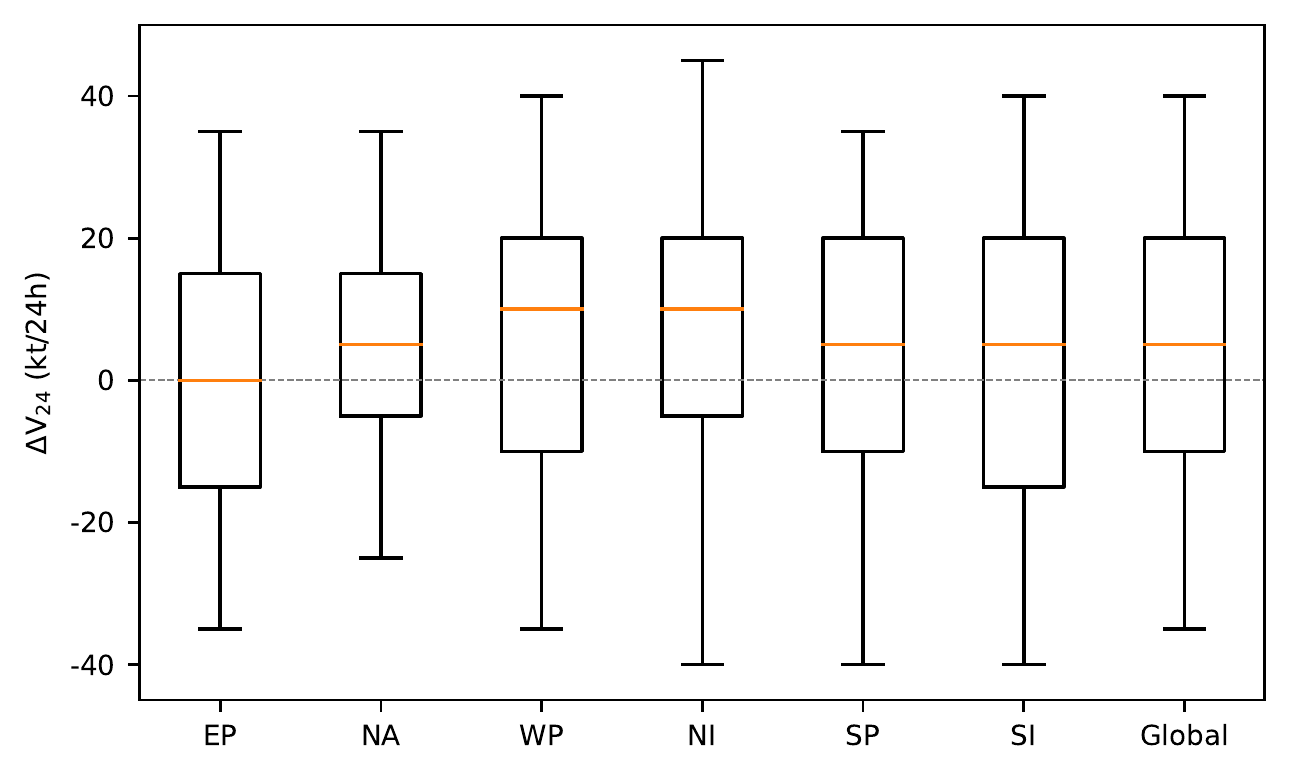}}
\caption{Boxplot of background $\Delta$V$_{24}$ for different basins. The lower and upper ends of the box show the 25$^\mathrm{th}$ and 75$^\mathrm{th}$ percentiles, the middle line shows the 50$^\mathrm{th}$ percentile, and the whiskers show the 5$^\mathrm{th}$ and 95$^\mathrm{th}$ percentiles, respectively.} \label{fig_vm_bp}
\end{figure}

Analyses for the individual basins were performed next. The global RI threshold of 45 kt/24h was found for most basins, except for the North Atlantic (NA) and East Pacific (EP), where the RI threshold was 40 kt/24h (Table~\ref{tab_ri}). 
These differences are likely to be an artifact since the measurements were provided by varied agencies.
However, it is also possible that the background $\Delta$V$_{24}$ varies among basins (Fig.~\ref{fig_vm_bp}), as already noticed by previous studies \cite<e.g.,>{xu_statistical_2015,xu_dependence_2018}.
For instance, The intensification rates are significantly higher in western North Pacific (WP) and North Indian Ocean (NI) than in EP and NA.
The 95$^{\mathrm{th}}$ percentile of $\Delta$V$_{24}$ in the former two basins was 40 and 40 kt/24h, respectively, while in the latter two basins it was 35 kt/24h.
% We added another Monte Carlo experiment, by adding \cite{bhatia_projected_2018}. 
Monte Carlo experiments also showed that the 95$^{\mathrm{th}}$ percentile of $\Delta$V$_{24}$ in EP and NA is 37.5 kt/24h, while for the other basins it is 41.5 kt/24h. 
Therefore different thresholds exist even the percentile-based method was used.
Such differences can be attributed to various environmental conditions.
Using an idealized numerical simulation,
\citeA{li_why_2021} noted a higher $\Delta$V$_{24}$ with a vertical sounding from WP than NA, especially when SST is lower than 28$^{\circ}$C.

Over all RI clusters, the initial intensity was approximately 60\textpm 15 kt, and the radius was approximately 45\textpm 20 km, while RMW was significantly larger (65\textpm 37 km) for the non-RI clusters (Table S2).
Moreover, intensity and size varied significantly among many basins.
For instance, the average $\Delta$V$_{24}$ of the RI cluster for the North Indian Ocean (60.4 kt/24h) was significantly higher than that for the North Atlantic (49.4 kt/24h).
The RI cases over the South Indian Ocean were initially significantly weaker and larger than those over the western North Pacific, whereas the typical $\Delta$V$_{24}$ and RI thresholds were similar for the two basins. 
Using a single threshold of 30 kt/24h failed to detect many of these differences. For example, a significant differences in RMW and initial intensity were observed between the RI events detected by clustering over the North Indian Ocean and western North Pacific. However, when the conventional threshold was used, there was no such difference in the inner-core size.
Significant difference (p$<$0.01) also exists in $\Delta$V$_{24}$ between Western and Eastern Pacific basins, as determined by the new threshold, while no difference could be detected (p$=$0.7) using the conventional threshold of 30 kt/24h.
However, we used a fixed period of 24h for all RI events and some of the events overlapped, and the results are likely to be different if consecutive RI events are considered as one.
% The initial 

\begin{table}
\caption{The RI thresholds, their corresponding percentile rank among the samples, and the mean and standard deviation of $\Delta$V$_{24}$, V$_{\mathrm{max}}$ and RMW of the RI clusters for the global and individual basins. (NATL, North Atlantic; EPAC, East Pacific; WPAC, Western North Pacific; SPAC, South Pacific; NIO, North Indian Ocean; SIO, South Indian Ocean). Note the percentiles were calculated using all $\Delta$V$_{24}$ instead of the positive values alone.}  \label{tab_ri}
\centering
\begin{tabular}{c m{7em} c c c c c}
\hline
  & RI threshold (kt/24h) & Percentile & $\Delta$V$_{24}$ (kt/24h) & V$_{\mathrm{max}}$ (kt) & RMW (km) \\
\hline
  Global & 45 & 97$^{\mathrm{th}}$ & 53.7~\textpm~9.8 & 61.7~\textpm~16.6 & 46~\textpm~20 \\
  NATL   & 40 & 98$^{\mathrm{th}}$ & 49.4~\textpm~11.0 & 64.3~\textpm~14.4 & 41~\textpm~23 \\
  EPAC & 40 & 96$^{\mathrm{th}}$ & 51.8~\textpm~9.9 & 59.9~\textpm~14.7 & 46~\textpm~19 \\
  WPAC   & 45 & 97$^{\mathrm{th}}$ & 55.7~\textpm~9.0 & 63.8~\textpm~15.7 & 44~\textpm~17 \\
  SPAC & 45 & 96$^{\mathrm{th}}$ & 55.3~\textpm~9.1 & 59.6~\textpm~15.6 & 51~\textpm~21 \\
  NIO &  45 & 96$^{\mathrm{th}}$ & 60.4~\textpm~10.0 & 56.5~\textpm~8.8 & 47~\textpm~19 \\
  SIO & 45 & 97$^{\mathrm{th}}$ & 55.0~\textpm~10.1 & 49.6~\textpm~11.3 & 53~\textpm~12 \\
\hline
% \multicolumn{5}{l}{}
\end{tabular}
\end{table}

\begin{figure}[ht!]
\centerline{\includegraphics[width=6in]{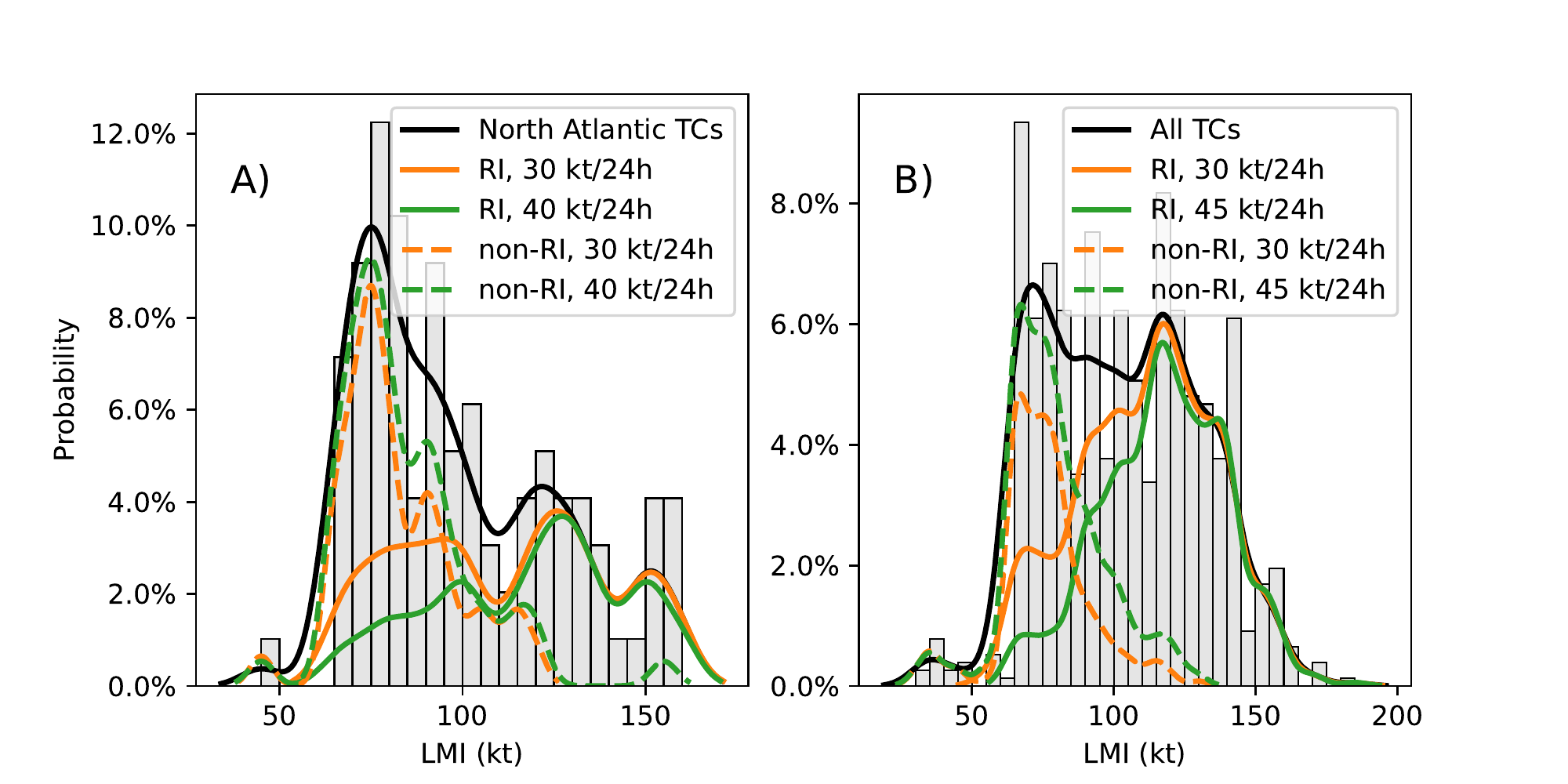}}
\caption{Distributions of LMI for (A) the western North Atlantic and (B) the global TCs. Probability Distribution Functions (PDFs) were calculated using 2004–2020 tropical cyclone LMI. The grey histograms and black line depict the distribution of all TCs. The orange and green lines show the PDFs for the TCs that underwent RI defined with by clustering (40 kt/24h for the North Atlantic and 45 kt/24h for the global TCs) and traditional thresholds (30 kt/24h), respectively. The PDFs were smoothed using kernel density estimation with a bandwidth of 0.2.} \label{fig_hist}
\end{figure}

The performance of the new threshold in explaining the distribution and variation of TC intensity was also examined.
\citeA{lee_rapid_2016} noted that the lifetime maximum intensity (LMI) follows a bimodal distribution, and the modes indicate two types of TCs: those that undergo RI (RI TCs) and those that do not (non-RI TCs). 
Here, we compared the effects of different thresholds on the distributions of the global and North Atlantic TCs (Fig.~\ref{fig_hist}).
Note only the TC data with both V$_{\mathrm{max}}$ and RMW recordings over the period 2004-2020 were used and thus the distributions differ slightly from \citeA{lee_rapid_2016}, who used all data from 1981-2012. 
Similar to their results, RI TCs, defined with either the clustering or traditional threshold, comprised the majority of the major TCs (LMI$\geq$96 kt, Category 3), with a peak at approximately 120 kt. However, the clustering threshold was more effective in separating the non-RI TCs, especially for the North Atlantic. When the clustering threshold was used, 80\% of RI TCs had an LMI of over 96 kt for the North Atlantic, and this ratio is 82\% for the global TCs. In contrast, with the threshold of 30 kt/24h, 65\% and 72\% of the RI TCs became major TCs for the North Atlantic and the globe, respectively.

\begin{figure}[ht!]
\centerline{\includegraphics[width=4in]{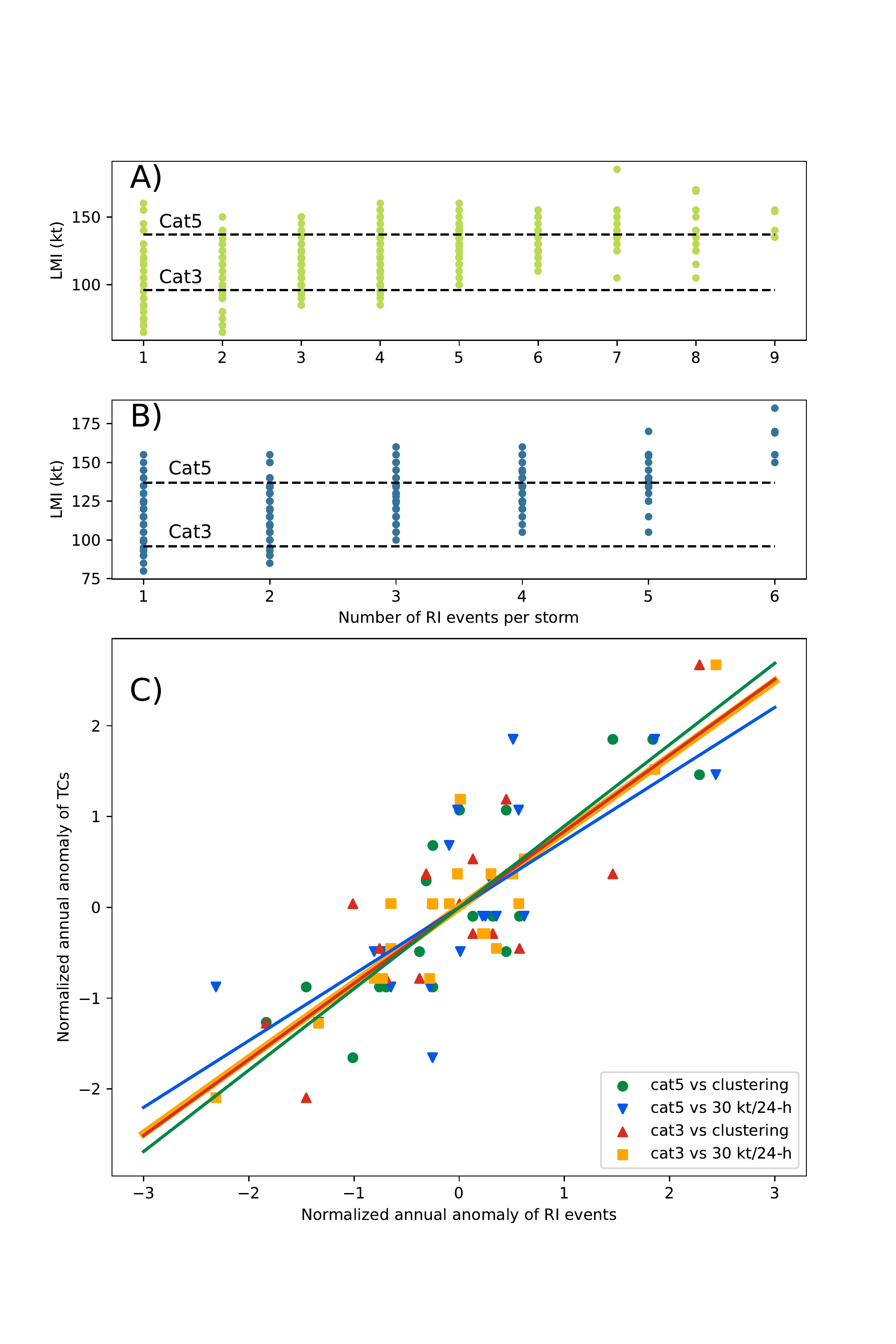}}
\caption{The distribution of LMI and the number of RI events per storm with (A) the threshold of 30 kt/24-h and (B) the clustering threshold. (C) Scatter plot of the normalized annual number of RI events versus the normalized annual number of TCs of different categories Each variable was normalized using its own mean and standard deviation prior to plotting. The solid lines show the linear regression. } \label{fig_ts}
\end{figure}

The high correlation between the number of RI events and major TCs further demonstrated the benefits of the new threshold. 
Fig.~\ref{fig_ts} plots the LMI against the number of RI events with the traditional and the proposed thresholds. When the clustering threshold was used, LMI clearly increased with the number of RI events per storm. However, when the threshold of 30 kt/24h was used, the number of RI events was not highly related to LMI, especially for the Category 5 TCs (LMI $\geq$ 137 kt). This was also demonstrated by a higher correlation between LMI and the number of RI events (0.62 vs 0.52, p$<$0.05). 
In addition, the number of RI events identified by the new threshold better explained the annual variation of the major TCs (Fig.~\ref{fig_ts}C), especially for Category 5 TCs, as indicated by relatively larger slops of their fitted lines. The correlations between the number of TCs stronger than Category 3 (LMI $\geq$ 96kt) and the clustering-based and conventional thresholds were both 0.81. However, the correlation between the number of Category 5 TCs and RI events with the new threshold was 10\% higher than that with the traditional one (0.85 vs 0.75, p$<$0.05).
These improvements demonstrated the potential of the proposed threshold in the research of major TCs. 

\begin{figure}[ht!]
\centerline{\includegraphics[width=6in]{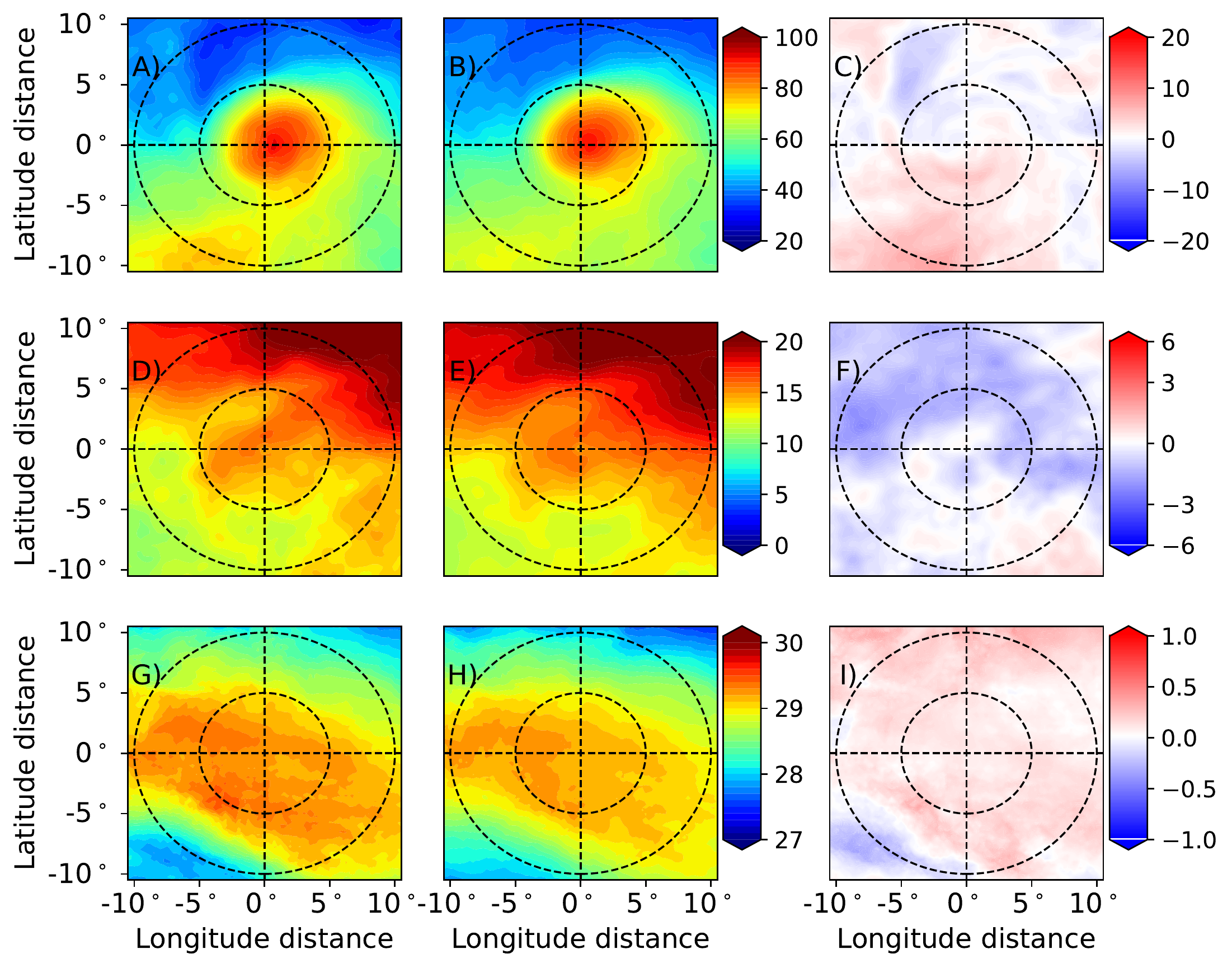}}
\caption{Composite fields of: (A), (B), (C) 600‐hPa relative humidity (units: \%; shaded); (D), (E), (F) 200- and 850-hPa vertical wind shear (units: m/s); and (G), (H), (I) sea surface temperature (units: $^{\circ}$C) for the (A), (D), (G) RI using the clustering threshold ($\geq$ 40 kt/24h); (B), (E), (H) RI using the traditional threshold ($\geq$ 30 kt/24h); and (C), (F), (I) the difference. The units for the x- and y-axes are degrees. The 2 circles indicate radii of 5 and 10 degrees from the TC center.} \label{fig_env}
\end{figure}

We further analyzed the environmental factors affecting the intensification processes in the western North Atlantic, namely mid-level relative humidity, deep vertical wind shear, and SST (Fig.~\ref{fig_env}). 
These variables were similar between the RI events, as defined by the clustering and traditional thresholds. Although a slightly higher mid-level humidity, lower vertical wind shear, and higher SST were observed for RI events defined by the clustering threshold, the differences are not statistically significant. The results partially validate our choice of using metrics representing the initial vortex property only.

% The sensitivity of clustering to varied combination of input variables and clustering algorithms are tested with the North Western Pacific TCs, where the intensity is stronger and sample is larger than other basins. Here the cluster number is set to 8 for consistency. When only the initial V$_{\mathrm{max}}$ is clustered with IR, K-Means successfully 

\section{Discussion and conclusion}
Rapid intensification of great importance in TC research.
Conventionally, RI is defined statistically.
% and physical processes were not considered during its definition.
% The conventional logic is to regard RI as an extreme event, and then study the processes related to it. 
In this study, we proposed a physically orientated threshold and incorporated properties of the vortex into its definition.
To do so, we treated RI as an extreme-detecting problem and cluster the initial V$_{\mathrm{max}}$ and RMW, with a subsequent $\Delta$V$_{24}$ for all global TCs from 2004 to 2020.
These variables were selected because they define the initial state of the vortex before RI, as demonstrated by previous theoretical and observational studies \cite<e.g.,>{wang_intensity-dependence_2021, li_why_2021, mallen_reexamining_2005, carrasco_influence_2014} and our sensitivity experiments. 
Over 12000 events extracted from the IBTrACS best-track data were clustered using the K-Means clustering algorithm; one cluster with a minimum $\Delta$V$_{24}$ of 45 kt/24h (23.2 m/s/24h) was thereby distinctly separated from the others. 
A significant gap between the RI and non-RI clusters was found, indicating that RI can be separated statistically and physically by vortex property.
The RI threshold was the same as the global value across all the individual basins except for the North Atlantic and East Pacific, where it was 40 kt/24h (20.6 m/s/24h). 

% Sensitivity tests were performed to compare the effects of clustering algorithms, the number of clusters, and, more importantly, the combination of input variables, including RMW, V$_{\mathrm{max}}$, TC fullness and R$_{34}$. 
% Monte Carlo experiments and an independent test for the 2021 hurricane season were also conducted to assess the impact of data uncertainty.
% The results suggested the threshold calculated by the clustering methods was robust. 
% The RI threshold was the same as the global value across all the individual basins except for the North Atlantic and East Pacific, where it was 40 kt/24h (20.6 m/s/24h). 
% In addition, although these variables showed larger inter-basin differences, the RI threshold itself appeared robust across the basins.

It should be noted that the results obtained from the clustering analysis rely heavily on data quality. 
RMW has a relatively large uncertainty because of limited temporal and spatial resolution in the satellite observations \cite{demuth_evaluation_2004}, although the data has been widely used for various basins \cite<e.g.,>{carrasco_influence_2014,xu_dependence_2018, xu_statistical_2015,wu_rapid_2021,li_how_2022}.
% For instance, the uncertainty for Atlantic hurricane outer size measurements is approximately 30\% \cite{Landsea2013AtlanticFormat}, which could undermine the definition produced in this study. 
Even for the Atlantic basin, where we have high confidence on the data quality, the uncertainty could be as high as 16 nm (30km) \cite{landsea_revised_2022}.
To assess the impact of data uncertainty, we conducted Monte Carlo experiments and an independent test for the 2021 hurricane season.
The results demonstrated the robustness of the proposed threshold, even with an uncertainty of 30 km in RMW. 
Nevertheless, this uncertainty is expected to decrease with improved observation facilities \cite<e.g.,>{combot_extensive_2020} and the clustering results could be revised with more accurate measurements, especially for the basins other than the Atlantic. 

% Although Monte Carlo experiments and independent test conducted in this work demonstrated the robustness of the proposed threshold, 
% To assess the potential influence of RMW uncertainty, we conducted Monte Carlo experiments and an independent test based on the data for the 2021 hurricane season.
% The results demonstrate the thresholds derived in this work is 

% which could undermine the definition produced in this study.
% Nevertheless, this uncertainty is expected to decrease with improved observation facilities \cite{Combot2020ExtensiveTrack} and the robustness of clustering would improve. 
% We also conducted Monte Carlo experiments to assess the potential influence of 

We do not compare performances of the two thresholds in operations due to the scope of this work.
The operational and practical choices of RI definition could be based purely on frequency. 
However, a more rigorous definition based on, for instance, the initial vortex state as advocated here may be desirable for case-based process study and trend analysis. 
As a first step, we present this new definition of RI from a different angle and emphasize on physical properties of the vortex behind the results. 
% As previously reported \cite{wang_intensity-dependence_2021,}, RI is associated with small RMW and medium initial intensity.
The clustering method identified an initial wind speed of 60\textpm15 kt and an RMW of 45\textpm20 km as the typical conditions for the RI cluster. 
While for the non-RI clusters, the average RMW mounts up to 65km.
Around these initial values, eye and eye-wall formation would be expected \cite{vigh_climatology_2012}. 
The onset wind speed cannot be much larger because, above this value, the frictional dissipation increases and $\Delta$V$_{24}$ is constrained as the TC approaches the maximum potential intensity \cite{wang_intensity-dependence_2021}. 
On the other hand, a small RMW reflects a high inner-core inertial stability and dynamical efficiency, which enhances intensification \cite{schubert_inertial_1982}. 
The inertial stability within the RMW was on the order of $10^{-3}$ s$^{-1}$ or about 20 f at 20 $^{\circ}$N and became important near these wind speeds.
A small RMW also favors intensification due to the conservation of angular momentum.
Both numerical and observational studies \cite{li_why_2021,li_how_2022,wang_outer_2020,sitkowski_intensity_2011,fischer_rapid_2020} showed the typical RMW value during RI is approximately 30 to 50 km, which is consistent with our results.
These results also suggest that accurate measurement and prediction of RMW near V$_{\mathrm{max}}$ of 60 kt would expected to play a significant role in improving RI predictions.

% It is also intuitive that the initial wind speed cannot be much larger because, above this value, $\Delta$V$_{24}$ is constrained as the TC approaches the maximum potential intensity. 
% Recently, \citeA{li_why_2021,li_how_2022} showed that a small RMW and rapid decrease of RMW are essential for RI due to the conservation of angular momentum.
% The typical RMW is approximately 30 to 50 km \cite{li_how_2022}, which is consistence with our results.

Moreover, an advantage of the clustering threshold is its ability to explain the bimodal distribution of the LMI. Both clustering and traditional thresholds can explain the secondary peak at around 120 kt. However, \citeA{lee_rapid_2016} reported that in the North Atlantic, 30\% RI TCs became minor TCs (LMI$<$96 kt). The ratio reduced to 20\% if the clustering threshold is used. Similar results were found for the global distribution. 
In addition, the new definition also showed a better correlation with the variation in the annual number of Category 5 TCs over the past decades compared to that of the traditional threshold, demonstrating the potential and importance of this new threshold for major TC research.

Environmental factors are undoubtedly important for TC intensification and many of such factors are included in the forecasting systems \cite<e.g.,>{kaplan_large-scale_2003}. However, previous studies found that the environmental conditions for the RI and intensifying TCs are similar \cite{hendricks_quantifying_2010}, and argued RI is a weak function of the environmental conditions \cite{kowch_are_2015}.
We also found that the difference of the mid-level humidity, vertical wind shear and SST between the RI events, with different thresholds, is marginal. 
% This indicates that RI is only weakly dependent on such factors if favorable environmental conditions are present, and validates our selection of using the inner-core metrics only. 
% Thus these factors are not sufficient, although might be essential, for 
Therefore favorable environmental factors are essential, but are likely insufficient for RI.
Although the environmental influences require further investigation, such results validate our selection of using the inner-core metrics only. 
% In addition, although the proposed threshold is more stringent than the conventional one, more studies are encouraged to examine the , especially for operational forecast since fewer TCs will meet the threshold. 

This work suggests the potential of clustering to define an RI threshold with an objective and plausible physical basis and demonstrates its advantages in explaining the distribution of major TCs. 
% In addition, this threshold is designed to identify RI cases that are dynamically different. 
We did not directly compare the operational applications of proposed and conventional thresholds, but the higher threshold proposed here would reduce the sample size of RI events, which may impact its operational usefulness in some aspects. 
Systematic analyses are thus required, including the ability of operational systems in detecting RI events, influences of different precursors, and damage caused by different RI events, before the proposed threshold is used in operation.

\acknowledgments
This study is sponsored by the National Natural Science Foundation of China (42006036, 42130409), National Key Research and Development Program (2017YFA0604202), Fundamental Research Funds for the Central Universities (B210202141) and Natural Environment Research Council/UKRI (NE/V017756/1). 

\section*{Open Research}
The data used in this study is the International Best Track Archive for Climate Stewardship (IBTrACS) Version 4 (v4r00) and publicly available at \url{https://www.ncdc.noaa.gov/ibtracs/index.php?name=ib-v4-access}.
ERA5 data is downloaded from \url{https://cds.climate.copernicus.eu/#!/home}.
% \section*{Open Research}
% The data used in this study is the International Best Track Archive for Climate Stewardship (IBTrACS) Version 4 (v4r00) and publicly available at \url{https://www.ncdc.noaa.gov/ibtracs/index.php?name=ib-v4-access}.

%% ------------------------------------------------------------------------ %%
%% References and Citations

%%%%%%%%%%%%%%%%%%%%%%%%%%%%%%%%%%%%%%%%%%%%%%%
%
% \bibliography{<name of your .bib file>} don't specify the file extension
%
% don't specify bibliographystyle
%%%%%%%%%%%%%%%%%%%%%%%%%%%%%%%%%%%%%%%%%%%%%%%

\bibliography{references.bib}

%Reference citation instructions and examples:
%
% Please use ONLY \cite and \citeA for reference citations.
% \cite for parenthetical references
% ...as shown in recent studies (Simpson et al., 2019)
% \citeA for in-text citations
% ...Simpson et al. (2019) have shown...
%
%
%...as shown by \citeA{jskilby}.
%...as shown by \citeA{lewin76}, \citeA{carson86}, \citeA{bartoldy02}, and \citeA{rinaldi03}.
%...has been shown \cite{jskilbye}.
%...has been shown \cite{lewin76,carson86,bartoldy02,rinaldi03}.
%... \cite <i.e.>[]{lewin76,carson86,bartoldy02,rinaldi03}.
%...has been shown by \cite <e.g.,>[and others]{lewin76}.
%
% apacite uses < > for prenotes and [ ] for postnotes
% DO NOT use other cite commands (e.g., \citet, \citep, \citeyear, \nocite, \citealp, etc.).
%

\end{document}

%% ------------------------------------------------------------------------ %%
%
%  TITLE
%
%% ------------------------------------------------------------------------ %%

%\includegraphics{agu_pubart-white_reduced.eps}

\title{Supporting Information for "Revisiting the definition of rapid intensification of tropical cyclones by clustering the initial intensity and inner-core size"}
%
% e.g., \title{Supporting Information for "Terrestrial ring current:
% Origin, formation, and decay $\alpha\beta\Gamma\Delta$"}
%
%DOI: 10.1002/%insert paper number here%

%% ------------------------------------------------------------------------ %%
%
%  AUTHORS AND AFFILIATIONS
%
%% ------------------------------------------------------------------------ %%

% List authors by first name or initial followed by last name and
% separated by commas. Use \affil{} to number affiliations, and
% \thanks{} for author notes.
% Additional author notes should be indicated with \thanks{} (for
% example, for current addresses).

% Example: \authors{A. B. Author\affil{1}\thanks{Current address, Antartica}, B. C. Author\affil{2,3}, and D. E.
% Author\affil{3,4}\thanks{Also funded by Monsanto.}}

\authors{Yi Li\affil{1,2,3}, Youmin Tang\affil{3,4}, Ralf Toumi\affil{5}, Shuai Wang\affil{5,6}}

\affiliation{1}{Key Laboratory of Marine Hazards Forecasting, Ministry of Natural Resources, Hohai University, Nanjing, China}
\affiliation{2}{Key Laboratory of Ministry of Education for Coastal Disaster and Protection, Hohai University, Nanjing, China}
\affiliation{3}{College of Oceanography, Hohai University, Nanjing, China}
\affiliation{4}{University of Northern British Columbia, Prince George, Canada}
\affiliation{5}{Department of Physics, Imperial College London, London, UK}
\affiliation{6}{Department of Civil and Environmental Engineering, Princeton University, Princeton, NJ, USA}
% \affiliation{1}{First Affiliation}
% \affiliation{2}{Second Affiliation}
% \affiliation{3}{Third Affiliation}
% \affiliation{4}{Fourth Affiliation}

%(repeat as many times as is necessary)

%% ------------------------------------------------------------------------ %%
%
%  BEGIN ARTICLE
%
%% ------------------------------------------------------------------------ %%

% The body of the article must start with a \begin{article} command
%
% \end{article} must follow the references section, before the figures
%  and tables.

\begin{article}

\clearpage

%Delete all unused file types below. Copy/paste for multiples of each file type as needed.
\noindent\textbf{Text S1. Configuration of sensitivity experiments}

Seven sensitivity experiments were performed to test different combinations of input variables.
The input variables include the initial intensity (V$_{\mathrm{max}}$), initial TC size metrics (RMW, R$_{34}$ and TC fullness) at t = 0 h, and the $\Delta$V$_{24}$ of the subsequent 24-h period.
The experiments are summarized in Table~\ref{tab_exp}.
In addition, we conducted a series of experiments using SOM to show the robustness of new definition regardless of clustering algorithm.

\clearpage
% the initial V$_{\mathrm{max}}$ and $\Delta$V$_{24}$ (hereafter V\_DV); initial V$_{\mathrm{max}}$, RMW and $\Delta$V$_{24}$ (hereafter V\_RMW\_DV); initial RMW and $\Delta$V$_{24}$ (hereafter RMW\_DV); initial V$_{\mathrm{max}}$, TC fullness [defined as $(1 - \frac{\mathrm{RMW}}{\mathrm{R}_{34}})$], and $\Delta$V$_{24}$ (hereafter V\_TCF\_DV); and initial V$_{\mathrm{max}}$, R$_{34}$, and $\Delta$V$_{24}$ (hereafter V\_R$_{34}$\_DV).

\noindent\textbf{Text S2. Results of sensitivity experiments}

\begin{figure}[ht!]
\centerline{\includegraphics[width=8in]{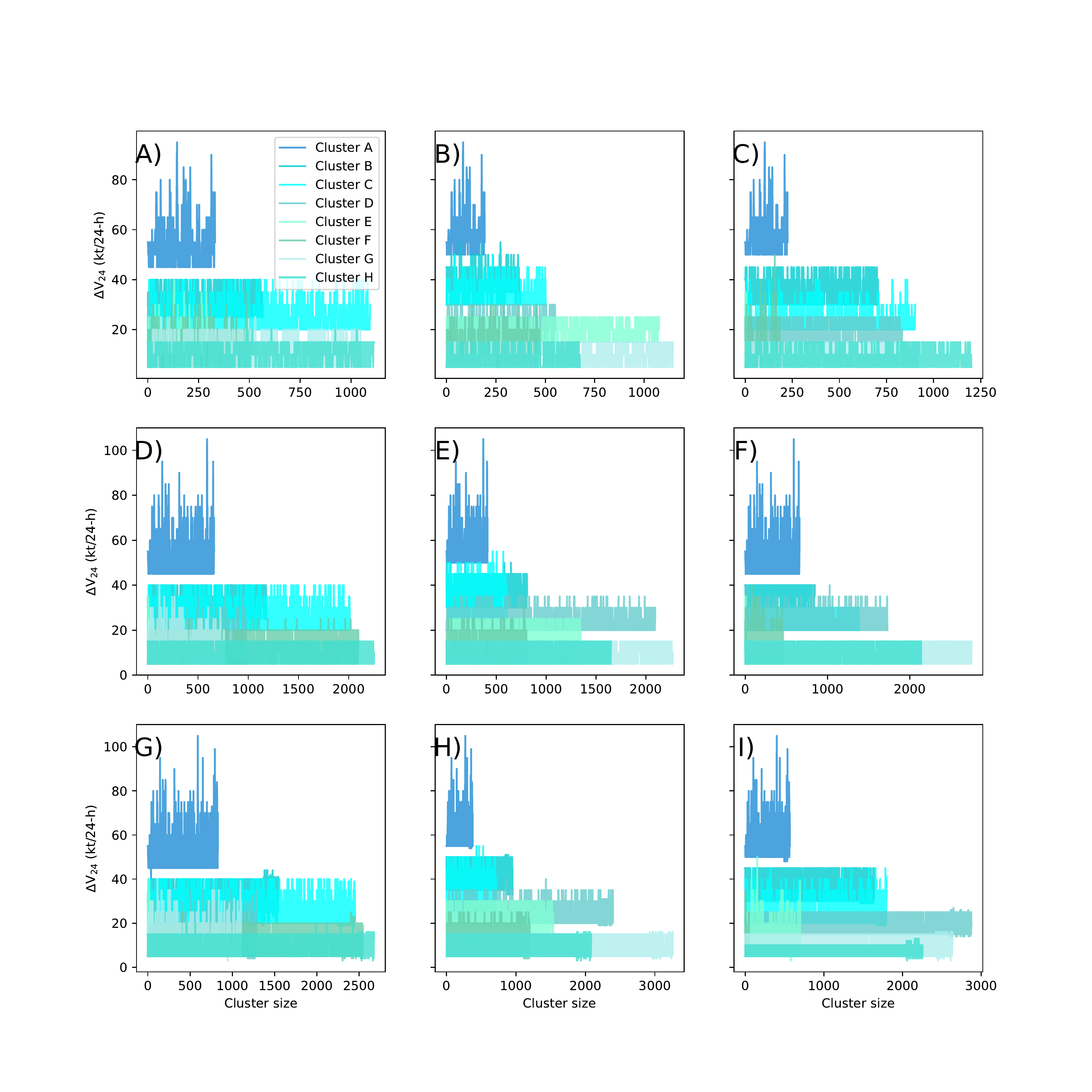}}
\caption{Clustering results of the (A) Exp\_V$_\mathrm{o}$\_R$_\mathrm{M}$, (B) Exp\_V$_\mathrm{o}$, and (C) Exp\_R$_\mathrm{M}$ experiments for the global basins with an input of 5000 TC events. (D), (E), and (F) are the same as (A), (B), and (C) but the input size was 10000. (G), (H), and (I) are the same as (A), (B), and (C) but the input size was 12530. The rapid intensification (RI) cluster is labeled as Cluster A in all subplots.} \label{fig_all_test}
\end{figure}

\begin{figure}[ht!]
\centerline{\includegraphics[width=8in]{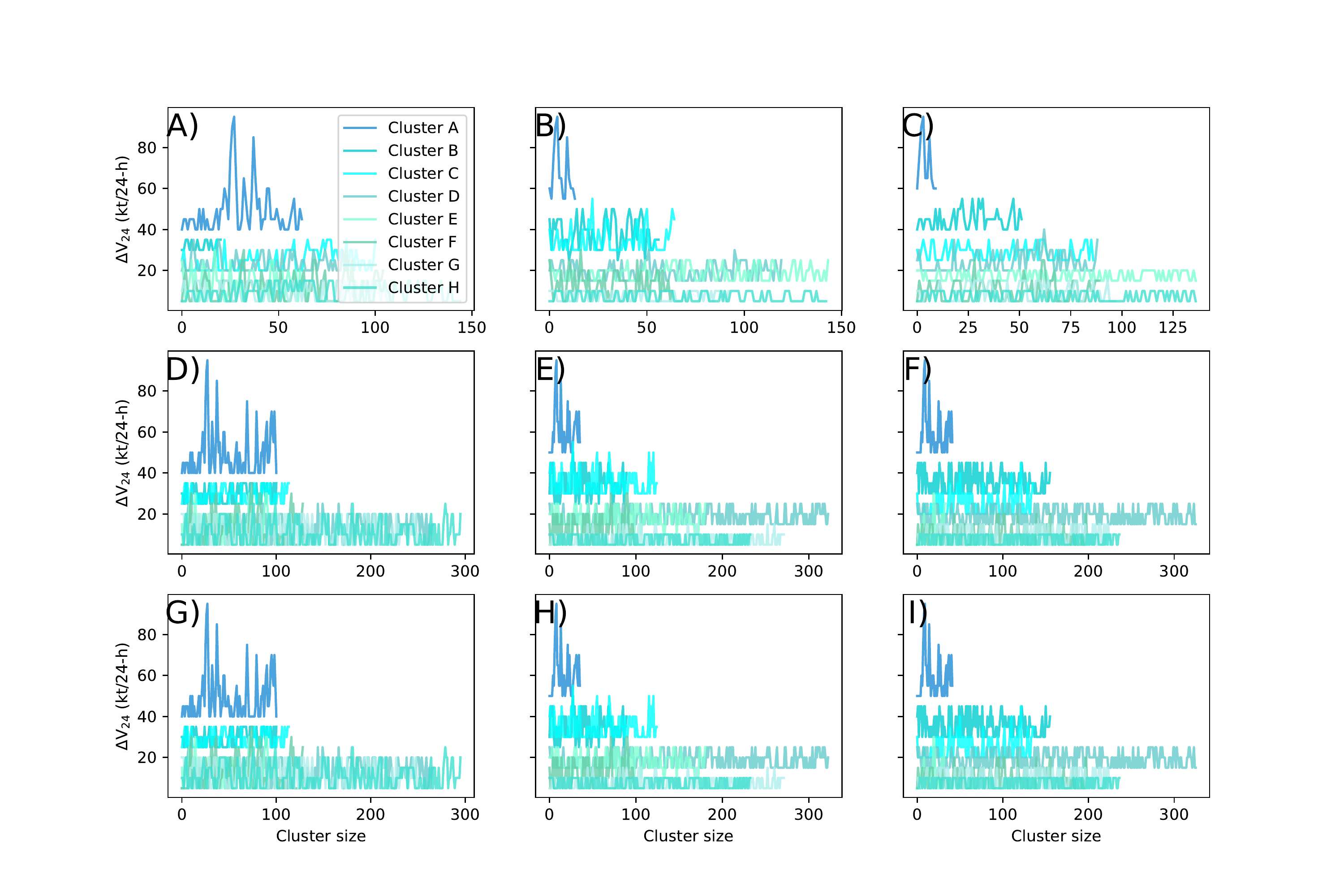}}
\caption{Same as Fig. S2, but for the North Atlantic basin.} \label{fig_na_test}
\end{figure}

\begin{figure}[ht!]
\centerline{\includegraphics[width=8in]{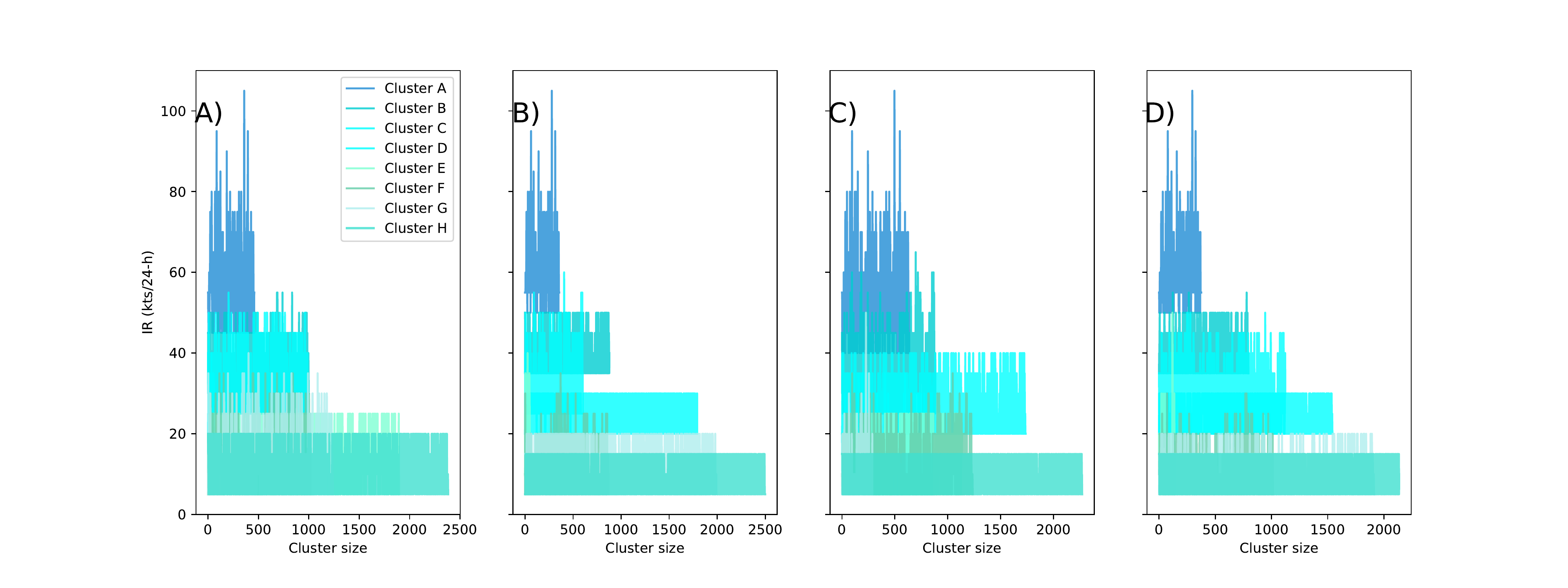}}
\caption{Clustering results of the (A) Exp\_V$_\mathrm{o}$\_R$_{34}$, (B) Exp\_R$_{34}$, (C) Exp\_TCF and (D) Exp\_Full experiments for the global basins.} \label{fig_test_tcf}
\end{figure}

The results of sensitivity experiments are described here in detail.
As reported in the main text, the optimal number of clusters ($K$) is 8 for Exp\_V$_\mathrm{o}$\_R$_M$, and results with 8 clusters are presented below for consistency. 
Exp\_V$_\mathrm{o}$ and Exp\_R$_\mathrm{M}$ showed that when either V$_{\mathrm{max}}$ or RMW alone was used with $\Delta$V$_{24}$, the clustering models still identified the RI group. 
For comparison, the results of Exp\_R$_{\mathrm{M}}$, Exp\_V$_{\mathrm{o}}$ and Exp\_V\_RMW are shown in Fig.~\ref{fig_all_test}.
Exp\_R$_{\mathrm{M}}$ outperformed Exp\_V$_{\mathrm{o}}$ because the overlap between RI and non-RI groups in $\Delta$V$_{24}$ was much smaller. 
However, Exp\_V$_\mathrm{o}$\_R$_M$ was preferable because of its larger cluster size (770 over 445 and 503) and robustness across different sample sizes. 
When the models were trained with 5000 or 10000 TC samples, only in Exp\_V$_\mathrm{o}$\_R$_M$ the model produced the same RI threshold (45 kt/24h) as training with the entire dataset of 12530 samples. 
In contrast, the thresholds in Exp\_R$_\mathrm{M}$ were 50 kt/24h at 5000 or 10000 samples, which increased to 55 kt/24h at 12530 samples. 
In addition, the gap between the RI and non-RI clusters was not obvious.
Similar results were found for Exp\_V$_\mathrm{o}$, and different thresholds were found when the sample size differed. 
A similar sensitivity test was performed for the North Atlantic, and the model with both RMW and initial V$_\mathrm{max}$ also produced the most robust clustering results (Fig.~\ref{fig_na_test}). 
Apart from RMW, other TC size metrics were also tested, including the average radius of the 34-kt wind speed (R$_{34}$), TC fullness and their combinations. 
However, we found such metrics were not efficient in separating RI clusters (Fig.~\ref{fig_test_tcf}). 
Additional experiments demonstrated that clustering with both positive and negative $\Delta$V$_{24}$ separated the weakening clusters but did not separate the RI and non-RI events as distinctly as with the positive $\Delta$V$_{24}$ alone. 
% The results of other sensitivity experiments showed that neither TC fullness nor R$_{34}$ generated an equivalent to the results of Exp\_V$_\mathrm{o}$\_R$_M$ (figure not shown). 
Therefore, the model with both RMW and initial V$_\mathrm{max}$ was selected, as stated in the main text. 

\noindent\textbf{Text S3. Results of the sensitivity experiments using the self organizing map (SOM) algorithm}

In addition, the self organizing map (SOM) method is used to test the sensitivity to clustering algorithms. 
% For consistency the number of clusters for SOM is set to 8. 
SOM also detects RI clusters with the threshold of 40 kt/24-h and the cluster size is approximately 1000 (Fig. S6), but the overlaps between RI and non-RI clusters are larger. Nonetheless, the results of SOM demonstrate the usefulness of clustering methods in defining RI and the robustness of the new threshold.
\end{article}

% Copy/paste for multiples of each file type as needed.

% enter figures and tables below here: %%%%%%%
%
%
%
%
% EXAMPLE FIGURES
% ---------------
% If you get an error about an unknown bounding box, try specifying the width and height of the figure with the natwidth and natheight options.
% \begin{figure}
%\setfigurenum{S1} %%You can change number for each figure if you want, not required. "S" prepended automatically.
% \noindent\includegraphics[natwidth=800px,natheight=600px]{samplefigure.eps}
%\caption{caption}
%\label{epsfiguresample}
%\end{figure}

% \begin{figure}[ht!]
% \centerline{\includegraphics[height=4in]{figs/IR_VO.pdf}}
% \caption{.} \label{fig_CDF}
% \end{figure}

% \begin{figure}[ht!]
% \centerline{\includegraphics[height=4in]{figs/kmeans_score.pdf}}
% \caption{} \label{fig_score}
% \end{figure} 
\clearpage

\begin{figure}[ht!]
\centerline{\includegraphics[width=8in]{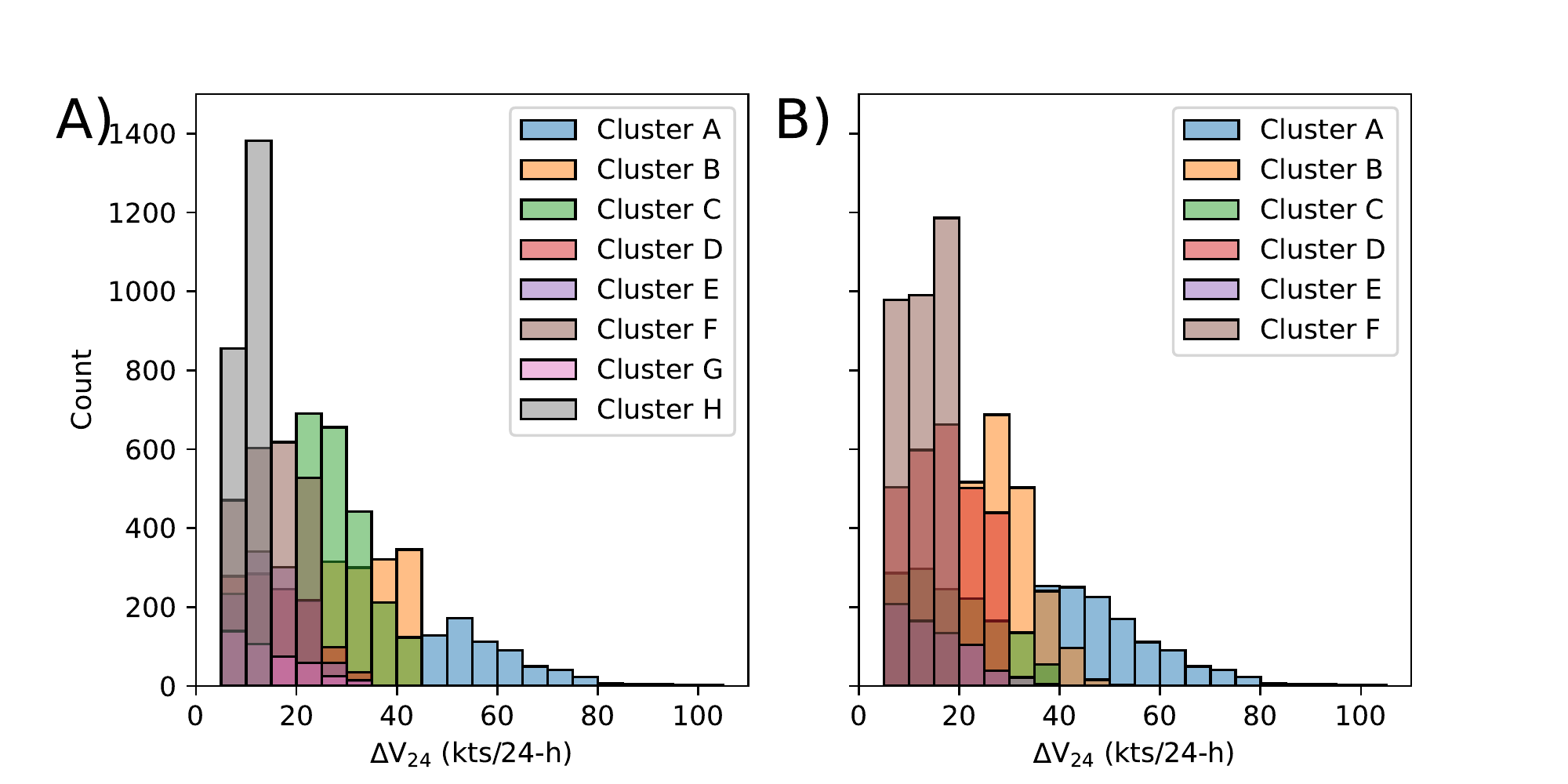}}
\caption{Same as Fig. 4, but with the results of all clusters plotted.} \label{fig_dv_vm_rm_8}
\end{figure} 

\clearpage

\clearpage

\clearpage

\clearpage

\begin{figure}[ht!]
\centerline{\includegraphics[width=8in]{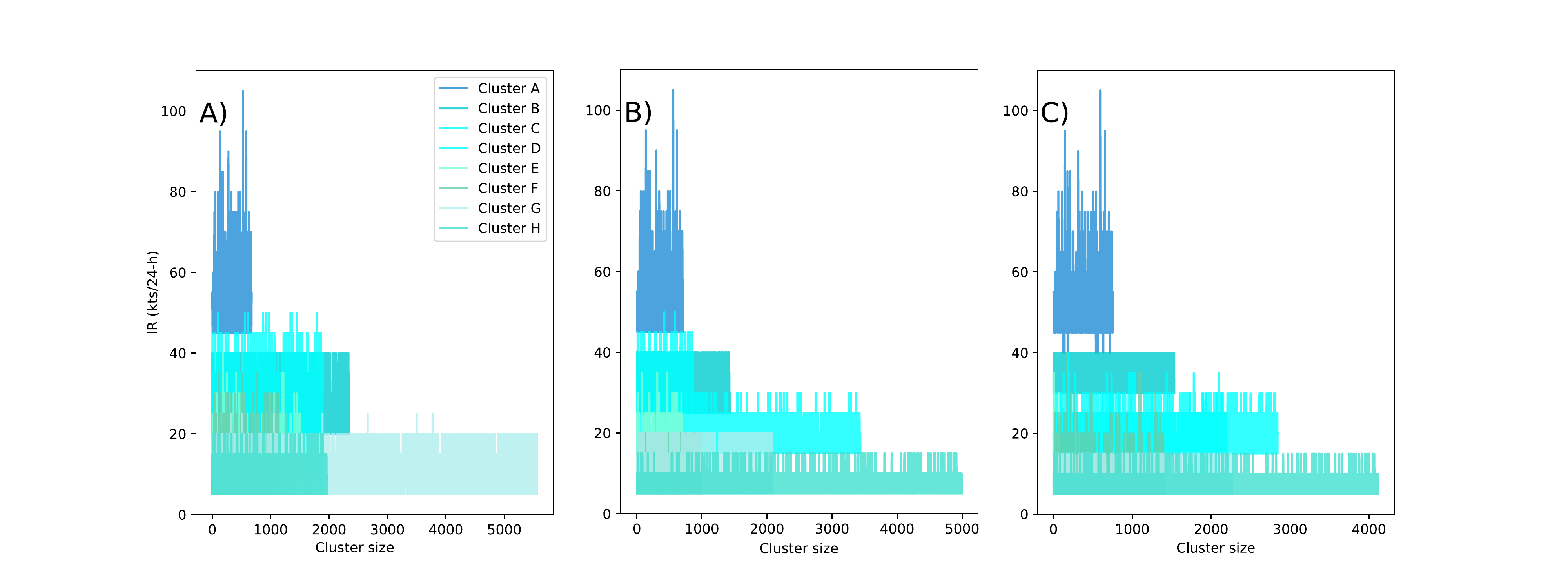}}
\caption{$\Delta$V$_{24}$ of different clusters with the input of (A) $\Delta$V$_{24}$, V$_\mathrm{max}$ and RMW; (B) $\Delta$V$_{24}$ and V$_\mathrm{max}$; and (C) $\Delta$V$_{24}$ and RMW for the global basins, when the TCs are clustered using the self-organizing map (SOM) method into 8 groups.} \label{fig_global_som}
\end{figure}

\clearpage
% \begin{figure}[ht!]
% \centerline{\includegraphics[height=4in]{figs/RI_Cat3.pdf}}
% \caption{The number of Major TCs (over Category 3) and RI TCs detected by different thresholds for each year over the period of 2001 to 2020.} \label{fig_ri_cat3}
% \end{figure}
%
%
% Giving latex a width will help it to scale the figure properly. A simple trick is to use \textwidth. Try this if large figures run off the side of the page.
% \begin{figure}
% \noindent\includegraphics[width=\textwidth]{anothersample.png}
%\caption{caption}
%\label{pngfiguresample}
%\end{figure}
%
%
%\begin{figure}
%\noindent\includegraphics[width=\textwidth]{athirdsample.pdf}
%\caption{A pdf test figure}
%\label{pdffiguresample}
%\end{figure}
%
% PDFLatex does not seem to be able to process EPS figures. You may want to try the epstopdf package.
%
%
% ---------------
% EXAMPLE TABLE
%

\begin{table}
\caption{The name and input variables of sensitivity experiments.}  \label{tab_exp}
\centering
\begin{tabular}{l l }
\hline
  Experiment Name &  Input Variables  \\
\hline
  Exp\_V$_\mathrm{o}$ & Initial V$_{\mathrm{max}}$, and $\Delta$V$_{24}$ \\
  Exp\_R$_\mathrm{M}$ & Initial RMW, and $\Delta$V$_{24}$ \\
  Exp\_R$_{34}$ & Initial R$_{34}$, and $\Delta$V$_{24}$ \\
  Exp\_TCF & Initial TC fullness, and $\Delta$V$_{24}$ \\
  Exp\_V$_\mathrm{o}$\_R$_\mathrm{M}$ & Initial V$_{\mathrm{max}}$, RMW, and $\Delta$V$_{24}$ \\
  Exp\_V$_\mathrm{o}$\_R$_{34}$ & Initial V$_{\mathrm{max}}$, R$_{34}$, and $\Delta$V$_{24}$ \\
  Exp\_Full & Initial V$_{\mathrm{max}}$, RMW, R$_{34}$, and $\Delta$V$_{24}$ \\
%   Exp\_V\_RM & Initial V$_{\mathrm{max}}$, RMW \\
%   Exp\_V\_TCF & Initial V$_{\mathrm{max}}$, RMW \\
\hline
% \multicolumn{5}{l}{}
\end{tabular}
\end{table}

\clearpage

\begin{table}
\caption{The RI thresholds, and the mean and standard deviation of $\Delta$V$_{24}$, V$_{\mathrm{max}}$ and RMW of the non-RI clusters for the global and individual basins. (NATL, North Atlantic; EPAC, East Pacific; WPAC, Western North Pacific; SPAC, South Pacific; NIO, North Indian Ocean; SIO, South Indian Ocean).}  \label{tab_nri}
\centering
\begin{tabular}{c m{7em} c c c c}
\hline
  & RI threshold (kt/24h) & $\Delta$V$_{24}$ (kt/24h) & V$_{\mathrm{max}}$ (kt) & RMW (km) \\
\hline
  Global & 45 & 17.1~\textpm~9.1 & 49.8~\textpm~25.4 &  65~\textpm~37 \\
  NATL   & 40 & 14.9~\textpm~8.6 & 50.5~\textpm~22.1 & 77~\textpm~42 \\
  EPAC & 40 & 16.3~\textpm~9.0 & 49.5~\textpm~23.7 & 65~\textpm~38 \\
  WPAC & 45 & 17.7~\textpm~9.7 & 50.3~\textpm~28.5 & 62~\textpm~31 \\
  SPAC & 45 & 17.8~\textpm~9.8 & 50.6~\textpm~24.8 & 59~\textpm~30 \\
  NIO & 45 & 17.2~\textpm~9.2 & 45.0~\textpm~23.9 & 65~\textpm~30 \\
  SIO & 45 & 17.2~\textpm~9.7 & 48.7~\textpm~24.7 & 60~\textpm~28 \\
\hline
% \multicolumn{5}{l}{}
\end{tabular}
\end{table}

\clearpage
% \begin{table}
% \settablenum{S1} %%Change number for each table
% \caption{Time of the Transition Between Phase 1 and Phase 2\tablenotemark{a}}
% \centering
% \begin{tabular}{l c}
% \hline
% Run  & Time (min)  \\
% \hline
%  $l1$  & 260   \\
%  $l2$  & 300   \\
%  $l3$  & 340   \\
%  $h1$  & 270   \\
%  $h2$  & 250   \\
%  $h3$  & 380   \\
%  $r1$  & 370   \\
%  $r2$  & 390   \\
% \hline
% \end{tabular}
% \tablenotetext{a}{Footnote text here.}
% \end{table}
% ---------------
%
% EXAMPLE LARGE TABLE (UPLOADED SEPARATELY)
%\begin{table}
%\settablenum{S1} %%Change number for each table
%\caption{Time of the Transition Between Phase 1 and Phase 2\tablenotemark{a}}
%\end{table}